\documentclass[11pt]{article}
\usepackage{amsmath}
\usepackage{amsfonts}
\usepackage{amsmath}
\usepackage{amsfonts}
\usepackage{graphicx}
\usepackage[usenames]{color}

\newcommand{\void}[1]{}

\numberwithin{equation}{section}
\begin{document}

\begin{titlepage}
\begin{center}
\begin{large}
\textbf{A Unified Approach to the Minimal Unitary Realizations of Noncompact Groups and  Supergroups}
\end{large}
\vspace{1cm}
\\
\begin{large}
Murat G\"{u}naydin$^{\dagger}$\footnote{murat@phys.psu.edu}
 and Oleksandr Pavlyk$^{\dagger,
 \ddagger}$\footnote{pavlyk@phys.psu.edu}
\end{large}
\\
\vspace{.35cm}
$^{\dagger}$ \emph{Physics Department \\
Pennsylvania State University\\
University Park, PA 16802, USA} \\
\vspace{.3cm}
and \\
\vspace{.3cm} $^{\ddagger}$ \emph{Wolfram Research Inc. \\100 Trade
Center Dr. \\ Champaign, IL 61820, USA} \\
\end{center}
\begin{abstract}
\noindent
 We study  the minimal unitary representations  of non-compact groups
and supergroups obtained by quantization of their geometric
realizations as quasi-conformal groups and supergroups. The
quasi-conformal groups $G$ leave generalized light-cones defined by
a quartic norm invariant and have maximal rank subgroups of the form
$H\times SL(2,\mathbb{R})$ such that $G/H\times SL(2,\mathbb{R})$
are para-quaternionic symmetric spaces. We give a unified
formulation of the minimal unitary representations of simple
non-compact groups  of type $A_2$, $G_2$, $D_4$,$F_4$, $E_6$, $E_7$,
$E_8$ and $Sp\left(2n,\mathbb{R}\right)$. The minimal unitary
representations of $Sp\left(2n,\mathbb{R}\right)$ are simply the
singleton representations and correspond to a degenerate limit of
the unified construction. The minimal unitary representations of the
other noncompact groups $SU\left(m,n\right)$, $SO\left(m,n\right)$ ,
$SO^*(2n)$ and $SL\left(m,\mathbb{R}\right)$ are also given
explicitly.

We extend our formalism to define and construct the corresponding
minimal representations of non-compact supergroups $G$ whose even
subgroups are of the form $H\times SL(2,\mathbb{R})$. If $H$ is
noncompact then the supergroup $G $ does not admit any unitary
representations, in general. The unified construction with $H$
simple or Abelian leads to the minimal representations of $G(3),
F(4)$ and $OSp\left(n|2,\mathbb{R}\right)$ (in the degenerate
limit). The minimal unitary representations of
$OSp\left(n|2,\mathbb{R}\right)$ with  even subgroups $SO(n)\times
SL(2,\mathbb{R})$ are the singleton representations. We also give
the minimal realization of the one parameter family of Lie
superalgebras $D\left(2,1;\sigma\right)$.
\end{abstract}
\end{titlepage}

\section{Introduction}

Inspired by the work on spectrum generating Lie algebras by physicists~\cite{spectgenalg}
Joseph introduced the concept of minimal unitary realizations of Lie algebras.
It is basically defined as a realization that exponentiates to a unitary
representation of the corresponding noncompact group on a Hilbert space
of functions depending on the minimal number of coordinates.  Joseph gave the
minimal realizations of the complex forms of classical Lie algebras
and of $G_2$ in a Cartan-Weil basis \cite{Jose74, Jose76}. The
existence of the minimal unitary representation of $E_{8(8)}$ within
the framework of Langland's classification was first proved by Vogan
\cite{Vogan81} .  Later, the minimal unitary representations of all
simply laced groups, were studied by Kazhdan and Savin\cite{KazSav90},
and Brylinski and Kostant
\cite{BryKos94,BryKos94a,BryKos95,BryKos96}. Gross and Wallach studied
the minimal representations of quaternionic real forms of exceptional
groups \cite{GroWal94}. For a review and the references on earlier
work on the subject in the mathematics literature prior to 2000 we
refer the reader to the review lectures of Jian-Shu Li~\cite{li2000}.

 The idea that the theta series of $\mathrm{E}_{8(8)}$
and its subgroups may describe the quantum supermembrane in various
dimensions~\cite{PW}, led Pioline, Kazhdan and Waldron~\cite{PKW} to
reformulate  the minimal unitary representations of simply laced
groups\cite{KazSav90}. In particular, they gave explicit
realizations of the simple root (Chevalley) generators in terms of
pseudo-differential operators for the simply laced exceptional
groups, together with the spherical vectors necessary for the
construction of modular forms.

Motivated mainly by the idea that the spectra of toroidally
compactified M/superstring theories  must fall into unitary
representations of their U-duality groups and towards the goal of
constructing these unitary representations G\"{u}naydin, Koepsell and
Nicolai   first studied the geometric realizations of U-duality
groups of the corresponding supergravity theories \cite{GKN1}. In
particular they gave geometric realizations of the U-duality groups
of maximal supergravity in four and three dimensions as conformal
and quasiconformal groups, respectively.  The  realization of the
3-dimensional U-duality group $E_{8(8)}$ of maximal supergravity
given in \cite{GKN1} as a quasiconformal group that leaves invariant
a generalized light-cone with respect to a quartic norm in 57
dimensions is the first known geometric realization of $E_8$.
 An $E_{7(7)}$ covariant construction of the minimal unitary
representation  of $E_{8(8)}$ by quantization of its geometric
realization as a quasi-conformal group \cite{GKN1} was then given in
\cite{GKN2}.
 The minimal unitary realization of the 3 dimensional U-duality group $E_{8(-24)}$ of
the exceptional supergravity \cite{GST} by quantization of its
geometric realization as a quasiconformal group was subsequently
given in \cite{GP1}. By consistent truncation the quasiconformal
realizations of the other noncompact exceptional groups can be
obtained from those of $E_{8(8)}$ and  $E_{8(-24)}$. Apart from
being the first known geometric realization of the exceptional group
of type $E_8$ the quasiconformal realization has some remarkable
features. First, there exist different real forms of all simple
groups that admit realizations as quasiconformal groups \footnote{
For $SU(1,1)=SL(2,\mathbb{R})=Sp(2,\mathbb{R})$ the quasiconformal
realization reduces to conformal realization.}. Therefore, the
quasiconformal realizations give a geometric meaning not only to the
exceptional groups that appear in the last row of the  famous Magic
Square \cite{MagicSquare} but also extend to certain real forms of
all simple groups.

Another remarkable property  of the quasiconformal realizations is
the above mentioned fact that their quantization leads, in a direct
and simple manner, to the minimal unitary representations of the
corresponding noncompact groups \cite{GKN2,GP1,GP2}.

Classification of the orbits of the actions of U-duality groups on
the BPS black hole solutions  in maximal supergravity and $N=2$
Maxwell-Einstein supergravity theories (MESGT) in five and four
dimensions given in \cite{fergun} suggested that four dimensional
U-duality may act as a spectrum generating conformal symmetry in
five dimensions \cite{fergun,GKN1}. Furthermore, the work of
\cite{GKN1} suggested that the 3 dimensional U duality group
$E_{8(8)}$ of maximal supergravity must similarly act as a spectrum
generating quasiconformal symmetry group in the charge space of BPS
black hole solutions in four dimensions extended by an extra
coordinate which was interpreted as black hole entropy. This extends
naturally to 3-dimensional U-duality groups of $N=2$ MESGTs acting
as spectrum generating quasiconformal symmetry groups in four
dimensions \cite{MG2005}. More recently it was conjectured that the
indexed degeneracies of certain $N=8$ and $N=4$ BPS black holes are
given by some automorphic forms related to the minimal unitary
representations of the corresponding 3 dimensional  U-duality groups
\cite{pioline2005}.

Motivated by the above mentioned  results and  conjectures
stationary and  spherically symmetric solutions of $ N\geq 2$
supergravities with symmetric scalar manifolds were recently studied
in \cite{GNPW}.  By utilizing the equivalence of four dimensional
attractor flow with the geodesic motion on  the scalar manifold of
the corresponding three dimensional theory  the authors of
\cite{GNPW} quantized the radial attractor flow, and argued that the
three-dimensional U-duality groups must act as spectrum generating
symmetry for BPS black hole degeneracies in 4 dimensions. They
furthermore suggested that these degeneracies may be related to
Fourier coefficients of certain modular forms of the 3-dimensional
U-duality groups, in particular those associated with their minimal
unitary representations.

The quasiconformal realizations of  noncompact groups represent
natural extensions of   generalized conformal realizations of some
of their subgroups and were studied from a spacetime point of view
in \cite{GP2}. The authors of \cite{GP2} studied in detail the
quasiconformal groups of generalized spacetimes defined by Jordan
algebras of degree three. The generic Jordan family of Euclidean
Jordan algebras of
 degree three describe extensions of the Minkowskian spacetimes
 by an extra ``dilatonic'' coordinate, whose rotation, Lorentz and
 conformal groups are $\mathrm{SO}(d-1), \mathrm{SO}(d-1,1)\times
 \mathrm{SO}(1,1)$ and $\mathrm{SO}(d,2)\times \mathrm{SO}(2,1)$,
 respectively. The generalized spacetimes described by simple Euclidean Jordan
 algebras of degree three correspond to extensions of Minkowskian
 spacetimes in the critical dimensions $(d=3,4,6,10)$ by a dilatonic
\emph{ and } extra ($2,4,8,16$) commuting \emph{spinorial}
coordinates, respectively. Their rotation, Lorentz and conformal
groups are those that occur in the first three rows of the Magic
Square \cite{MagicSquare}.
    For the generic Jordan
 family the quasiconformal groups are $\mathrm{SO}(d+2,4)$. On the
 other hand,
  the quasiconformal groups of spacetimes defined by simple Euclidean Jordan algebras
  of degree are  $\mathrm{F}_{4(4)}$, $\mathrm{E}_{6(2)}$,
 $\mathrm{E}_{7(-5)}$ and $\mathrm{E}_{8(-24)}$. The conformal
 subgroups of these quasiconformal groups are $Sp(6,\mathbb{R}),
 SU^*(6), SO^*(12)$ and $E_{7(-25)}$, respectively.

In this paper we give a unified construction of the minimal unitary
representations of noncompact groups by quantization of their
geometric realizations as quasiconformal groups and extend it to the
construction of the minimal representations of noncompact
supergroups. In section 2 we explain the connection between minimal
unitary representations of noncompact groups $G$ and  their unique
para-quaternionic symmetric spaces of the form $G/H\times
SL(2,\mathbb{R})$, which was used in \cite{bg97} to give a
classification and minimal realizations of the real forms of
infinite dimensional nonlinear quasi-superconformal Lie algebras
that contain the Virasoro algebra as a subalgebra \cite{fradline}.
In section 3, using some of the results of \cite{bg97}  we give a
unified construction of the minimal unitary representations of
simple noncompact groups with $H$ simple or Abelian. A degenerate
limit of the unified construction leads to the minimal unitary
representations of  the symplectic groups $Sp(2n,\mathbb{R})$, which
is discussed in section 4. In sections 5, 6 ,7 and 8 we give the
minimal unitary realizations of $SO(p+2,q+2)$, $SO^*(2n+4)$,
$SU(n+1,m+1)$ and $SL(n+2,\mathbb{R})$, respectively. In section 9
we extend our construction to the minimal  realizations of
noncompact supergroups and present the unified construction of the
minimal representations of supergroups whose even subgroups are of
the form $H\times SL(2,\mathbb{R})$ with $H$ simple.The construction
of the minimal unitary realizations of $OSp(N\vert2,\mathbb{R})$
corresponds to a degenerate limit of the unified construction and is
discussed in section 10, where we also give the minimal realization
of $D(2,1;\alpha)$. Preliminary results of sections 3 and 9 appeared
in \cite{sashathesis}.

\section{Minimal Unitary Representations of Noncompact Groups and
Para-Quaternionic Symmetric Spaces}

 The minimal dimensions for simple
non-compact groups  were determined by Joseph \cite{Jose74,Jose76}.
For a particular  noncompact group $G$ the minimal dimension $\ell$
can be found by considering the 5-graded decomposition of its Lie
algebra $\mathfrak{g}$, determined by a distinguished
$\mathfrak{sl}(2,\mathbb{R})$ subalgebra, of the form
\begin{equation}
 \mathfrak{g}= \mathfrak{g}^{-2}  \oplus  \mathfrak{g}^{-1}  \oplus
  \left(  \mathfrak{g}^{0}  \oplus \Delta \right)
  \oplus \mathfrak{g}^{+1}  \oplus  \mathfrak{g}^{+2}
\end{equation}
where $\mathfrak{g}^{\pm 2}$ are 1-dimensional subspaces each, and
$\Delta$ is the dilatation generator that determines the five
grading. The generators belonging to the subspace
 $\mathfrak{g}^{-2} \oplus \Delta \oplus \mathfrak{g}^{+2}$
form the $\mathfrak{sl}(2\,, \mathbb{R} )$ subalgebra in question.
The minimal dimension $\ell$ is simply
\begin{equation}
   \ell=\frac{1}{2} \dim \left(\mathfrak{g}^{+1}\right) +1
\end{equation}
If we denote the subgroup generated by the grade zero subalgebra
$\mathfrak{g}^0$ as $H$, then the quotient
\begin{equation}
 \frac{G}{H\times SL(2,\mathbb{R})}
\end{equation}
is a para-quaternionic symmetric space in the terminology of
\cite{alexcort}. Our goal in this paper is to complete the
construction of the minimal unitary representations of all such
non-compact groups by quantization of their quasiconformal
realizations. Remarkably, the  para-quaternionic symmetric spaces
arose earlier in the classification \cite{bg97} of infinite
dimensional nonlinear quasi-superconformal Lie algebras that contain
the Virasoro algebra as a subalgebra \footnote{These infinite
dimensional non-linear  algebras were proposed as symmetry algebras
that unify perturbative (Virasoro) and non-perturbative (U-duality)
symmetries  \cite{mgcoralgables}.}. Below we list all simple
noncompact groups $G$ of this type and their subgroups $H$
\cite{bg97}:

\begin{center}
\begin{tabular}{|c|c|}
\hline $G$ & $H$ \\ \hline
$SU(m,n)$ & $U(m-1,n-1)$ \\ \hline
$SL(n,\mathbb{R})$ & $GL(n-2,\mathbb{R})$ \\ \hline
$SO(n,m)$ & $SO(n-2,m-2)\times SU(1,1)$ \\ \hline
$SO^*(2n)$ & $SO^*(2 n-4)\times SU(2)$ \\ \hline
$Sp(2n,\mathbb{R})$ & $Sp(2n-2,\mathbb{R})$ \\ \hline
$E_{6(6)}$ & $SL(6,\mathbb{R})$ \\ \hline
$E_{6(2)}$ & $SU(3,3)$ \\ \hline
$E_{6(-14)}$ & $SU(5,1)$ \\ \hline
$E_{7(7)}$ & $SO(6,6)$ \\ \hline
$E_{7(-5)} $ & $SO^*(12)$ \\ \hline
$E_{7(-25)}$ & $SO(10,2) $ \\ \hline
$E_{8(8)}$ & $E_{7(7)}$ \\ \hline
$E_{8(-24)} $ & $E_{7(-25)} $ \\ \hline
$F_{4(4)}$ & $Sp(6,\mathbb{R})$ \\ \hline
$G_{2(2)}$ & $SU(1,1)$ \\ \hline
\end{tabular}
\end{center}
The minimal unitary representations of the
exceptional groups ($F_4$, $E_6$, $E_7$, $E_8$) and of $SO(n,4)$ as well
as the corresponding quasiconformal realizations were given in
\cite{GKN2,GP1,GP2}.

\section{Unified Construction of the minimal unitary realizations
of non-compact groups with $ H $ simple or Abelian.}
Consider the 5-graded decomposition of the Lie algebra
$\mathfrak{g}$ of $G$
\begin{equation}
 \mathfrak{g}^{-2}  \oplus  \mathfrak{g}^{-1}  \oplus \left(  \mathfrak{g}^{0}  \oplus \Delta \right)
  \oplus  \mathfrak{g}^{+1}  \oplus  \mathfrak{g}^{+2} \nonumber
\end{equation}
Let $J^a$ denote generators of the Lie algebra $\mathfrak{g}^0$ of
$H$
\begin{subequations}\label{eq:alg}
\begin{equation}
   \left[ J^a \,, J^b \right] = {f^{ab}}_c J^c
\end{equation}
where $a,b,...=1,...D$ and let $\rho$ denote the symplectic
representation by which $\mathfrak{g}^0$ acts on $\mathfrak{g}^{\pm
1}$
\begin{equation}
   \left[ J^a \,, E^{\alpha} \right] = {\left(\lambda^{a}\right)^\alpha}_\beta E^\beta
\qquad
   \left[ J^a \,, F^{\alpha} \right] =  {\left(\lambda^{a}\right)^\alpha}_\beta F^\beta
\end{equation}
where $E^\alpha$, $ \alpha, \beta, ..= 1,..,N= \dim (\rho)$ are
generators that span the subspace $\mathfrak{g}^{-1}$
\begin{equation}
   \left[ E^\alpha \,, E^\beta  \right] = 2 \Omega^{\alpha\beta} E
\end{equation}
and $F^\alpha$ are generators that span $\mathfrak{g}^{+1}$
\begin{equation}
   \left[ F^\alpha \,, F^\beta  \right] = 2 \Omega^{\alpha\beta} F
\end{equation}
and $\Omega^{\alpha\beta}$ is the symplectic invariant ``metric'' of
the representation $\rho$.  The negative  grade generators form a
Heisenberg subalgebra since
\begin{equation}
  \left[E^{\alpha}, E \right] = 0
\end{equation}
with the grade -2 generator $E$  acting as its central charge.
Similarly the positive grade generators form a Heisenberg algebra
with the grade +2 generator $F$ acting as its central charge. The
remaining nonvanishing commutation relations of $g$ are
\begin{equation}
\begin{aligned}
  F^\alpha &= \left[ E^\alpha \,, F \right] \cr
  E^\alpha &= \left[ E \,, F^\alpha \right] \cr
  \left[E^{\alpha} , F^{\beta}\right] &= - \Omega^{\alpha\beta} \Delta + \epsilon \lambda_a^{\alpha\beta} J^a
\end{aligned}
\qquad
\quad
\begin{aligned}
\left[\Delta, E^{\alpha} \right] &= - E^{\alpha} \cr
\left[\Delta, F^{\alpha} \right] &= F^{\alpha} \cr
\left[\Delta, E\right] &= -2 E \cr
\left[\Delta, F \right] &= 2 F
\end{aligned}
\end{equation}
\end{subequations}
where $\Delta$ is the generator that determines the five grading and
$\epsilon$ is a  parameter to be determined.

 We shall realize the generators using bosonic oscillators
$\xi^\alpha$ satisfying the canonical commutation relations
\begin{equation}
   \left[ \xi^\alpha \,, \xi^\beta \right] = \Omega^{\alpha\beta}
\end{equation}
The grade -1, -2 generators and those of $H$ can be realized easily
as
\begin{equation}
  E = \frac{1}{2} y^2 \qquad E^\alpha = y \,\xi^\alpha \qquad
 J^a =  - \frac{1}{2}  {\lambda^a}_{\alpha\beta} \xi^\alpha \xi^\beta
\end{equation}
where $y$, at this point, is an extra ``coordinate'' such that
$\frac{1}{2}y^2$ acts as the central charge of the Heisenberg
algebra formed by the negative grade generators.

Now there may exist different real forms  of $ G $  with different
subgroups $ H $.  For reasons that will become obvious  we shall
assume that a real form  of $ G $ exists for which $ H $  is simple.
We shall  follow the conventions of \cite{bg97} throughout this
paper except for the  occasional use of Cartan labeling  of simple
Lie algebras whenever we are not considering  specific real forms.

The quadratic Casimir operator of the Lie algebra $\mathfrak{g}^0$
of $H$ is
\begin{equation}
 \mathcal{C}_2\left(\mathfrak{g}^0\right) = \eta_{ab} J^a J^b
\end{equation}
where $\eta_{ab}$ is the Killing metric of $H$.  The minimal
realizations given in \cite{GKN2,GP1,GP2} and the results of
\cite{bg97} suggest  an Ansatz
 for the grade +2 generator $F$  of the form
\begin{equation} \label{eq:exprF}
   F = \frac{1}{2} p^2 + \kappa y^{-2} \left( \mathcal{C}_2 + \mathfrak{C} \right)
\end{equation}
where $p$ is the momentum conjugate to the coordinate $y$
\begin{equation}
[y,p]=i
\end{equation}
and $\kappa$ and $\mathfrak{C}$ are some constants to be
determined later. This implies then
\begin{eqnarray}
  F^\alpha &=& \left[E^{\alpha}, F\right]= i p \, \xi^\alpha + \kappa y^{-1} \left[ \xi^\alpha \,, \mathcal{C}_2
  \right]  \nonumber \\
 &= & i p \, \xi^\alpha -\kappa y^{-1} \left[ 2 \, {\left(\lambda^{a}\right)^{\alpha}}_\beta \xi^\beta J_a
  + \, C_\rho \,\xi^\alpha \right]
\end{eqnarray}
where $C_\rho$ is the eigenvalue of the second order Casimir of $H$
in the representation $\rho$.\footnote{ Note that the indices
$\alpha, \beta,..$ are raised and lowered with the antisymmetric
symplectic metric $\Omega^{\alpha\beta}=-\Omega^{\beta\alpha}$ that
satisfies
$\Omega^{\alpha\beta}\Omega_{\gamma\beta}=\delta^{\alpha}_{\beta} $
and $V^\alpha = \Omega^{\alpha\beta} V_\beta $, and $V_\alpha
=V^\beta \Omega_{\beta\alpha} $. In particular, we have $V^\alpha
W_\alpha = - V_\alpha W^\alpha $.}

We choose the normalization of the representation matrices $\lambda$
as in \cite{fradline,bg97}
\begin{equation}
 \lambda^{a,\alpha \beta} \lambda_{a~\delta}^{~\gamma}
  - \lambda^{a,\gamma \alpha} \lambda_{a~\delta}^{~\beta}
  = - \frac{C_{\rho}}{N+1}  \left(\Omega^{\alpha \beta}
  \delta^{\gamma}_{~\delta}
  - 2 \Omega^{\beta \gamma} \delta^{\alpha}_{~\delta}
  + \Omega^{\gamma \alpha} \delta^{\beta}_{~\delta}\right), \label{eq:necsuff}
\end{equation}
which implies

\begin{equation}
\lambda^a_{\alpha\beta}
\lambda_a^{\beta\gamma}= -C_{\rho} \delta^{\gamma}_{\alpha}
\end{equation}
The unknown constants in our Ansatz will be determined by requiring
that generators satisfy the commutation relations \eqref{eq:alg} of
the Lie algebra $\mathfrak{g}$. We first consider commutators of
elements of $\mathfrak{g}^1$ and $\mathfrak{g}^{-1}$

\begin{equation}
\label{eq:EF}
   \left[ E^\alpha \,, F^\beta \right] = i \left( y \, p \right) \Omega^{\alpha\beta}
     - \xi^\beta \, \xi^\alpha + \kappa \left[ \xi^\alpha \,, \left[ \xi^\beta \,,\mathcal{C}_2 \right] \right]
\end{equation}
which, upon using the identity,
\begin{equation}
  \left[ \xi^\alpha \,, \mathcal{C}_2 \right] = - 2 \, {\left(\lambda^{a}\right)^{\alpha}}_\beta \xi^\beta J_a - \, C_\rho \, \xi^\alpha
  \label{eq:aux1}
\end{equation}
leads to
\begin{equation}
  \left[ E^\alpha \,, F^\beta \right] = - \Delta \Omega^{\alpha\beta} + \left\{
         \frac{3 \kappa C_\rho}{1 + N} - \frac{1}{2} \right\} \left(  \xi^\alpha \xi^\beta + \xi^\beta \xi^\alpha \right) - 6 \kappa \left(\lambda^a\right)^{\alpha\beta} J_a
\end{equation}
where $\Delta = - \frac{i}{2} \left( yp+py\right)$. Now the
bilinears  $\left(  \xi^\alpha \xi^\beta + \xi^\beta \xi^\alpha
\right) $ generate the Lie algebra of $\mathfrak{c}_{N/2}$ $\left
(\mathfrak{sp}\left(N\right) \right )$ under commutation. Hence for
those Lie algebras $\mathfrak{g}$ whose subalgebras $\mathfrak{g}^0$
are different from $\mathfrak{c}_{N/2}$ closure requires that the
coefficient of the second term vanish
\begin{equation}
\label{eq:kappa1}
   \frac{3 \kappa C_\rho}{1 + N } - \frac{1}{2} = 0
\end{equation}

For Lie algebras $\mathfrak{g}$ whose subalgebras $\mathfrak{g^0}$
are of type $\mathfrak{c}_{N/2}$ we have
\begin{equation*}
  \left(\lambda_a\right)^{\alpha\beta} J^a \approx \xi^\alpha \xi^\beta +
      \xi^\beta \xi^\alpha
\end{equation*}
Hence we do not get any constraints  on $\kappa $  from the above
commutation relation.

Next, let us compute  the commutator
\begin{equation}
\label{eq:FF}
   \left[ F^\alpha \,, F^\beta \right] =
   \frac{\kappa}{y^{2}} \left(   - \xi^\alpha \left[ \xi^\beta \,, \mathcal{C}_2 \right] +
          \xi^\beta \left[ \xi^\alpha \,, \mathcal{C}_2 \right] +
          \kappa \left[ \left[ \xi^\alpha \,, \mathcal{C}_2 \right]  ,
                        \left[ \xi^\beta \,, \mathcal{C}_2 \right]
                 \right]
   \right) - p^2 \Omega^{\alpha\beta}
\end{equation}
Using \eqref{eq:necsuff}, \eqref{eq:aux1} we write the terms linear
in $\kappa$ on the right hand side as

\begin{multline}
   \frac{\kappa}{y^{2}}\left( - \xi^\alpha \left[ \xi^\beta \,, \mathcal{C}_2 \right] +
          \xi^\beta \left[ \xi^\alpha \,, \mathcal{C}_2 \right] \right)=\\
    \frac{\kappa}{y^{2}} \left(   C_\rho \Omega^{\alpha\beta} +2 \left( \xi^\alpha {\left( \lambda^a\right)^\beta}_\gamma -
     \xi^\beta {\left( \lambda^a\right)^\alpha}_\gamma \right) \xi^\gamma
     J_a \right).
\end{multline}
The terms  quadratic in $\kappa$ on the right hand side gives
\begin{eqnarray*}
  \kappa \left[ \left[ \xi^\alpha \,, \mathcal{C}_2 \right]  ,
                        \left[ \xi^\beta \,, \mathcal{C}_2 \right]
                 \right] =
     \frac{12 \kappa C_\rho}{N+1} \left( \xi^\alpha {\left( \lambda^a\right)^\beta}_\gamma -
     \xi^\beta {\left( \lambda^a\right)^\alpha}_\gamma \right) \xi^\gamma J_a -
               \kappa C_\rho^2 \Omega^{\alpha\beta} \nonumber\\
    + 4 \kappa \left(
         3 \left( \lambda^b \lambda^a \right)^{\alpha\beta} J_a J_b -
                   2 \left( \lambda^b \lambda^a \right)^{\beta\alpha} J_a J_b  +
                  {f_{ab}}^c {\left(\lambda^a \right)^\alpha}_\mu
                                        {\left(\lambda^b \right)^\beta}_\nu \xi^\mu\xi^\nu J_c
           \right)
\end{eqnarray*}
Using these two expressions above  \eqref{eq:FF} becomes
\begin{equation}
\begin{split}
&\left[ F^\alpha \,, F^\beta \right] = - 2 \left(  \frac{1}{2} p^2 + \frac{1}{y^2}
     \left( \frac{\kappa^2}{2} C_\rho^2 - \frac{\kappa}{2} C_\rho \right)  \right) \Omega^{\alpha\beta} \\
    &+ \frac{4 \kappa}{y^2} \left( \xi^\alpha {\left( \lambda^a\right)^\beta}_\gamma -
     \xi^\beta {\left( \lambda^a\right)^\alpha}_\gamma \right) \xi^\gamma J_a \\
    & + \frac{4 \kappa^2}{y^2} \left(
         3 \left( \lambda^b \lambda^a \right)^{\alpha\beta} J_a J_b -
                   2 \left( \lambda^b \lambda^a \right)^{\beta\alpha} J_a J_b  +
                  {f_{ab}}^c {\left(\lambda^a \right)^\alpha}_\mu
                                        {\left(\lambda^b \right)^\beta}_\nu \xi^\mu\xi^\nu J_c
           \right)
\end{split}
\label{eq:g1g1=g2}
\end{equation}
Now the right hand side of \eqref{eq:g1g1=g2} must equal
$2 \Omega^{\alpha\beta} F $ with
\begin{equation*}
  F= \frac{1}{2} p^2 + \kappa y^{-2} \left( \mathcal{C}_2 +
\mathfrak{C} \right)
\end{equation*}
per our Ansatz.
Contracting the right hand side of \eqref{eq:g1g1=g2} with
$\Omega_{\beta\alpha}$ we get
\begin{equation}
  -N \left( p^2 + \frac{1}{y^2}\left(\kappa^2 C_\rho^2 - \kappa C_\rho \right)\right)
   - \frac{1}{y^2} \kappa \left( - 16 + 20 \kappa i_\rho \ell^2
    - 4 \kappa C_\text{adj} \right) \mathcal{C}_2
\end{equation}
where $i_\rho$ is the Dynkin index of the representation $\rho$ of
$H$ and $C_{adj}$ is the eigenvalue of the second order Casimir in
the adjoint of $H$. To obtain this result one uses the fact that
\begin{equation*}
\lambda^a_{\alpha\beta} \lambda^{b,\alpha\beta} = -i_{\rho} \ell^2
\eta^{ab}
\end{equation*}
where  $\ell$ is the length of the longest
root of $H$.\footnote{The length squared $\ell^2$ of the longest
root is normalized such that it is 2 for the simply laced algebras,
4 for $B_n, C_n$ and $F_4$ and 6 for $G_2$. The $i_\rho, C_\rho$ and
$\ell$ are related by $ i_\rho= \frac{NC_\rho}{D\ell^2}$ where
$D=\dim(H)=\dim(g^0)$.}
Using
\begin{equation}
C_{adj}= - \ell^2 h^\vee
\end{equation}
where $h^\vee$ is the dual Coxeter number of $\mathfrak{g}^0$
subalgebra of $\mathfrak{g}$, the closure then requires
\begin{equation}\label{eq:kappa2}
  \left( - 8 + 10 \kappa i_\rho \ell^2
    + 2 \kappa h^\vee \ell^2 \right) =  N
\end{equation}
   Equations \eqref{eq:kappa1} and
\eqref{eq:kappa2}, combined with
\begin{equation}
    i_\rho \ell^2 = \frac{N}{D} C_\rho
\end{equation}
imply
\begin{equation}\label{eq:restr1}
   \frac{h^\vee}{i_\rho} = \frac{3 D}{N (N+1)} (N+8) -5
\end{equation}

The validity of the above expression can be verified explicitly  by
comparing with Table 1 of \cite{bg97}, relevant part of which is
collected in Table \ref{tbl:bg97}  for convenience.
\begin{table}
\centering
\begin{tabular}{|l|l|l|l|l|} \hline
$g^0$ & $D$ & $h^\vee$ & $N=dim\rho$ & $i_{\rho}$ \\  \hline
$\mathfrak{c}_n$ & $n(2n+1)$ & $n+1$ & $2n$ & $\frac{1}{2}$\\
$\mathfrak{a}_5$ & 35 & 6 & 20 & 3\\
$\mathfrak{d}_6$ & 66 & 10 & 32 & 4\\
$\mathfrak{e}_{7}$ & 133 & 18 & 56 & 6\\
$\mathfrak{c}_3$ & 21 & 4 & 14 & $\frac{5}{2}$\\
$\mathfrak{a}_1$ & 3 & 2 & 4 & 5 \\ \hline
  \end{tabular}
\caption{%
The list of grade zero subalgebras $g^0$ with dual Coxeter number $h^\vee$
that are simple and with  irreducible action $\rho$ on grade $+1$
subspace. $i_{\rho}$ is the Dynkin index of the representation $\rho$. \label{tbl:bg97}%
}%
\end{table}
Furthermore, it was shown in \cite{bg97} that all the groups and the
corresponding symplectic representations listed in the above table
satisfy the equation
\begin{equation}
h^\vee = 2 i_\rho \left ( \frac{D}{N} + \frac{3D}{N(1+N)} -1 \right)
\end{equation}
which was obtained as a consistency condition for the existence of
certain class of infinite dimensional nonlinear quasi-superconformal
algebras. Comparing this equation with the equation \eqref{eq:restr1}  we see
that they are consistent with each other  if
\begin{equation} \label{eq:magic}
   D = \frac{3 N \left(N+1\right) }{N+16}
\end{equation}

\noindent Requirement of $\left[ F \,, F^\alpha \right] = 0$ leads to the condition
\begin{equation}
\label{eq:NoGrade3Condition}
   \xi^\alpha \left( \mathcal{C}_2 + \mathfrak{C} \right) +
   \left( \mathcal{C}_2 + \mathfrak{C} \right) \xi^\alpha +
   \kappa \left[ \mathcal{C}_2 \,, \left[ \xi^\alpha \,, \mathcal{C}_2 \right] \right] = 0
\end{equation}
Using \eqref{eq:aux1} and $\left[ \mathcal{C}_2 \,, J^a \right] = 0$ we arrive at
\begin{equation}\label{eq:FFalpha}
\begin{split}
  2\, \xi^\alpha \left( \mathcal{C}_2 + \mathfrak{C} \right) +
  2 \left( 1 - \kappa C_\rho \right) C_\rho \xi^\alpha +
  2 \left( 1 - \kappa C_\rho \right) {\left(\lambda^a\right)^\alpha}_\beta \, \xi^\beta J_a & \\
  - 4 \, \kappa {\left( \lambda^a \lambda^b \right)^\alpha}_\beta \,& \xi^\beta J_b J_a = 0
\end{split}
\end{equation}
In order to extract restrictions on $\mathfrak{g}$ implied by the above equation we contract it
with $\xi^\gamma \Omega_{\gamma \alpha}$ and obtain
\begin{equation}\label{eq:restr2}
  \frac{h^\vee}{i_\rho} = \frac{D}{N (N+1)} (N-8) + 1 \,.
\end{equation}
It agrees with \eqref{eq:restr1} provided \eqref{eq:magic} holds true.
Making use of \begin{equation*} N = 2 (g^\vee -2) \end{equation*} where
$g^\vee$ is the dual Coxeter number of the Lie algebra
$\mathfrak{g}$ and  \eqref{eq:magic} we obtain
\begin{equation}\label{eq:magic2}
   \dim\left( \mathfrak{g} \right) = 2 + 2 N + 1 + \dim \left( \mathfrak{g}^0 \right) =
   1 + 2 \left( N+1\right) + D =  2 \frac{\left(g^\vee + 1 \right) \,
    \left(5 g^\vee - 6 \right) }{ g^\vee + 6}
\end{equation}

Equation \eqref{eq:NoGrade3Condition} and the requirement of the right hand side of \eqref{eq:g1g1=g2}
 to equal to $2 \Omega^{\alpha\beta} F$ imply restriction on matrices $\lambda^a$ for
  which \eqref{eq:restr1} and \eqref{eq:restr2} are only necessary conditions.
  We expect  these conditions to be derivable from the identities satisfied by
  the corresponding
   Freudenthal triple systems \cite{WorkInProgress} that underlie the
   quasiconformal actions and the  minimal realizations \cite{GKN1, GP2}.

\begin{table}\centering
\begin{tabular}{l|l|l|l}
  $\mathfrak{g}$ & $\dim(\mathfrak{g})$ & $g^\vee$ & Eqtn. \eqref{eq:magic2} holds ?\cr
  \hline
  $\mathfrak{a}_n$ & $n^2 + 2 n$ & $n + 1$ & for  $\mathfrak{a}_1$
  and
  $\mathfrak{a}_2$ only\cr
  $\mathfrak{b}_n$ & $2 n^2 + n$ & $2 n - 1$ & no \cr
  $\mathfrak{c}_n$ & $2 n^2 + n$ & $n + 1$ & no\cr
  $\mathfrak{d}_n$ & $2 n^2 - n$ & $2 n - 2$ & for $\mathfrak{d}_4$ only\cr
  $\mathfrak{e}_6$ & $78$ & $12$ & yes \cr
  $\mathfrak{e}_7$ & $133$ & $18$ & yes \cr
  $\mathfrak{e}_8$ & $248$ & $30$ & yes \cr
  $\mathfrak{f}_4$ & $52$ & $9$ & yes \cr
  $\mathfrak{g}_2$ & $14$ & $4$ & yes
\end{tabular}
\caption{Dimensions and dual Coxeter numbers of simple  Lie
algebras. In order for the Lie algebra to admit a non-trivial
5-graded decomposition its dimension must be greater than 6. This
rules out $\mathfrak{sl}(2)$ for which \eqref{eq:magic2} also holds.
\label{tab:data}}
\end{table}

By going through the list of  simple Lie algebras \cite{Gilmore}
collected for convenience in Table \ref{tab:data} we see that the
equation \eqref{eq:magic} is valid only for the Lie algebras of
simple groups $A_1, A_2,$ $ G_2,  D_4,  F_4, $$ E_6,  E_7$ and
$E_8$. For $A_1$ our realization reduces simply to the conformal
realization.  With the exception of $D_4$,
 what these groups have in common is the fact
 that their subgroups  $H$ are either simple or
 one dimensional Abelian as expected  by the consistency with our Ansatz.
 The reason our Ansatz also covers the case of $D_4$ has to do with
 its unique properties. The subalgebra $\mathfrak{g}^0$ of $\mathfrak{d}_4$
is the direct sum of three copies of $\mathfrak{a}_1$
\begin{equation*}
  \mathfrak{g}^0 =\mathfrak{a}_1 \oplus \mathfrak{a}_1 \oplus
  \mathfrak{a}_1
\end{equation*}
The eigenvalues of the quadratic Casimirs of these subalgebras
$\mathfrak{a}_1$ as well as their Dynkin indices in the representation
$\rho$ coincide as required by the consistency with our Ansatz. These
groups appear in the last row of the so-called Magic Triangle
\cite{MagicTriangle} which extends the Magic Square of Freudenthal,
Rozenfeld and Tits \cite{MagicSquare}.

There is, in addition, an infinite family of non-compact groups for which
$H$ is simple, namely the  noncompact symplectic groups
$\mathrm{Sp}(2n+2,\mathbb{R})$ with $\dim{\rho}=2n$ and $H=
Sp(2n,\mathbb{R})$. However, as remarked above,  the constraint
\eqref{eq:kappa1} and hence the equation \eqref{eq:magic} do not
follow from our Ansatz for the symplectic groups. The quartic
invariant becomes degenerate for symplectic groups and  the minimal
unitary realizations  reduce to free boson construction of the
singleton representations for these groups  as will be discussed in
the next section.

The minimal unitary realizations of noncompact groups appearing in
the Magic Triangle \cite{MagicSquare,MagicTriangle} can be obtained
by consistent truncation of the minimal unitary realizations of the
groups appearing in its last row \cite{GP1,GKN1,GP2}.  We should
stress that there are different real forms of the groups appearing
in the Magic Square or its straightforward extension to the Magic
Triangle. Different real forms in our unified construction
correspond to different hermiticity conditions on the bosonic
oscillators $\xi^{\alpha}$ \cite{GP1}. After specifying the
hermiticity properties of the oscillators $\xi^{\alpha}$ one goes to
a Hermitian (anti-hermitian) basis of the Lie algebra $\mathfrak{g}$
with purely imaginary (real) structure constants to calculate the
Killing metric which determines the real form corresponding to the
minimal unitary realization.

The quadratic Casimir operator of the Lie algebra constructed in a
unified manner above is given by
\begin{equation}
  \mathcal{C}_2 \left( \mathfrak{g} \right) = J^a J_a + \frac{2 \, C_\rho}{N+1} \left( \frac{1}{2}\, \Delta^2 + E F + F E \right) - \frac{ C_\rho}{N+1}
  \Omega_{\alpha\beta} \left( E^\alpha F^\beta + F^\beta E^\alpha \right)
\end{equation}
which, upon using \eqref{eq:kappa1} and the following identities
\begin{equation}
\begin{split}
   \frac{1}{2}\, \Delta^2 + E F + F E &= \kappa\left( J^a J_a + \mathfrak{C} \right) - \frac{3}{8} \\
   \Omega_{\alpha\beta}\left( E^\alpha F^\beta + F^\beta E^\alpha \right) &= 8 \, \kappa J^a J_a + \frac{N}{2} + \kappa C_\rho N
\end{split}
\end{equation}
 that follow from our Ansatz,  reduces to  a c-number
\begin{equation}
\begin{split}
\mathcal{C}_2 \left( \mathfrak{g} \right) &=  \mathfrak{C} \left( \frac{8 \kappa C_\rho}{N+1} - 1 \right) - \frac{3}{4} \, \frac{C_\rho}{N+1} - \frac{N}{2} \, \frac{C_\rho}{N+1} - \frac{\kappa C_\rho^2 N}{N+1} \\
  &=_\text{(using eq.\eqref{eq:kappa1})} - \frac{C_\rho}{36} \, \frac{ (N + 4)\,(5\, N + 8)}{N+1}
\end{split}
\end{equation}
as required by irreducibility. We should note that this result
agrees with explicit calculations for the Lie algebras of the Magic
Square  in  \cite{GP1}. In the normalization chosen there $\kappa =
2$ and hence $ 12 C_\rho = N+1$. Then, using $N=2 g^\vee - 4$ we get
\begin{equation}
    \mathcal{C}_2 \left( \mathfrak{g} \right) = - \frac{1}{108}  \left( 5 g^\vee - 6 \right) g^\vee.
\end{equation}

\section{The minimal unitary  representations of \\ $Sp \left(2n+2, \mathbb{R}\right)$}
The Lie algebra of $Sp\left(2n+2, \mathbb{R}\right)$  has a
5-grading of the form
\begin{equation}
 \mathfrak{sp}\left(2n+2,\mathbb{R}\right) = E \oplus E^\alpha \oplus
   \left( \mathfrak{sp}\left(2n, \mathbb{R}\right) \oplus \Delta \right) \oplus F^\alpha \oplus F
\end{equation}
where $E^\alpha = y \xi^\alpha$, and $E = \frac{1}{2} y^2$.
Generators of the grade zero subalgebra $\mathfrak{g}^0 =
\mathfrak{sp}\left(2n, \mathbb{R}\right)$ are given simply by the
symmetrized bilinears ( modulo normalization)
\begin{equation}
  -2 \left(\lambda_a\right)^{\alpha\beta} J^a = \xi^\alpha \xi^\beta +
      \xi^\beta \xi^\alpha
\end{equation}
which is simply the singleton ( metaplectic) realization of
$\mathfrak{sp}(2n, \mathbb{R})$.
 The quadratic Casimir of $\mathfrak{sp}\left(2n,
\mathbb{R}\right)$ in the singleton realizaton is simply a
\emph{c}-number. As stated in the previous section the constraint
equation \eqref{eq:kappa1} that follow from the commutation
relations $[E^{\alpha}, F^{\beta}]$ can not be imposed in the case
of symplectic Lie algebras $Sp\left(2n+2, \mathbb{R}\right)$.
However  the equation \eqref{eq:NoGrade3Condition} that follows from
our Ansatz requires that $\mathcal{C}_2 + \mathfrak{C}= 0$ or
$\kappa =0$ for the symplectic Lie algebras $Sp\left(2n+2,
\mathbb{R}\right)$. In other words $F = \frac{1}{2} p^2$. Thus
\begin{eqnarray}
  F^\alpha &=& \left[ E^\alpha, F \right] = i p \xi^\alpha \\
  && \left[E^\alpha, F^\beta\right] = i \left(y p \right) \Omega^{\alpha\beta} - \xi^\beta \xi^\alpha
\nonumber \\
&=& -\frac{i}{2} \Delta - \frac{1}{2} (\xi^\alpha \xi^\beta +
\xi^\beta \xi^\alpha )
\end{eqnarray}
 Thus the minimal unitary realization of the symplectic group $Sp(2n+2,\mathbb{R})$ obtained by quantization
 of its quasiconformal realization is simply  the singleton realization in terms bilinears
 of the $2n+2$ oscillators (annihilation and creation
operators) $\left( \xi^\alpha, y, p\right)$. The quadratic Casimir
of $\mathfrak{sp}\left(2n+2, \mathbb{R}\right)$ is also a
\emph{c}-number.

That the quadratic Casimir is a \emph{c}-number is only a necessary
requirement for the irreducibility of the corresponding
represenation.  For the above singleton realization the entire Fock
space of all the oscillators decompose into the direct sum of the
two  inequivalent singleton representations that have  the same
eigenvalue of the quadratic Casimir. They are both unitary lowest
weight representations. By choosing a definite polarization one can
define $n+1$  annihilation operators
\[ a_0= \frac{1}{\sqrt{2}} (y+ip) \]
\[a_i =\frac{1}{\sqrt{2}} (\xi^i +i \xi^{n+i}) \qquad i=1,2,..,n\]
and $n+1$ creation operators
\[ a^0= \frac{1}{\sqrt{2}} (y-ip) \] \[
a^i =\frac{1}{\sqrt{2}} (\xi^i -i \xi^{n+i}) \] in terms of the
(n+1) coordinates and (n+1) momenta. The vacuum vector $ |0\rangle$
annihilated by all the annihilation operators
\[ a_0|0\rangle = a_i|0\rangle=0 \] is the lowest weight vector of
the  ``scalar'' singleton irrep of $Sp(2n+2,\mathbb{R})$ and  (n+1)
vectors
\[ a^0|0\rangle, a^i |0\rangle \]
form the lowest K-vector of the other singleton irrep of
$Sp(2n+2,\mathbb{R})$. In other words the lowest K vector of the
scalar singleton is an $SU(n+1)$ scalar, while the lowest K-vector
of the other singleton irrep is a vector of $SU(n+1)$ subgroup of
$Sp(2n+2,\mathbb{R})$. Both lowest K-vectors carry a nonzero $U(1)$
charge.

The reason for the reduction of the minimal unitary realizations of
the Lie algebras of  symplectic groups $Sp(2n+2,\mathbb{R})$ to
bilinears, and hence to a free boson construction, is the fact that
there do not exist any nontrivial quartic invariant of
$Sp(2n,\mathbb{R})$  defined by an irreducible symmetric tensor in
the fundamental representation $2n$. We have only the skew symmetric
symplectic invariant tensor $\Omega{\alpha\beta}$  in the
fundamental representation, which when contracted with $\xi^{\alpha}
\xi^{\beta}$ gives a c-number.

In the light of the above results one may wonder how the
quasi-conformal realization of $\mathfrak{sp}\left(2n+2,
\mathbb{R}\right)$ can be made manifest.  Before quantisation we
have $2n+1$ coordinates $\mathbb{X} = \left(X^\alpha, x\right)$ on
which we realize $\mathfrak{sp}\left(2n+2, \mathbb{R}\right)$:
\begin{equation}
  \mathcal{N} \left( \mathbb{X}\right) = I_4\left(X^\alpha\right)-x^2
\end{equation}
since $I_4\left(X^\alpha\right) = 0$. With the ``twisted'' difference
 vector defined as \cite{GKN1}
\begin{equation}
 \delta\left( \mathbb{X}, \mathbb{Y} \right) = \left( X^\alpha - Y^\alpha, x-y + \left< X, Y \right> \right)
\end{equation}
The equation defining the generalized lightcone  \[ \mathcal{N}
\left( \delta\left(\mathbb{X}, \mathbb{Y} \right)\right) =0 \] then
reduces to
\begin{equation}
   x - y + \left< X, Y \right> = 0
\end{equation}
where $\left< X, Y \right>= \Omega_{\alpha\beta} X^\alpha Y^\beta$.
By reinterpreting the coordinates $(X^\alpha, x)$ and $(Y^\alpha ,
y)$ as projective coordinates in 2n+2 dimensional space
\begin{equation*}
 x= \frac{\xi^0}{\xi^{n+1}}
\end{equation*}
\begin{equation*}
 X^\alpha = \frac{\xi^\alpha}{\xi^{n+1}}
\end{equation*}
\begin{equation*}
  y= \frac{\eta^0}{\eta^{n+1}}
\end{equation*}
\begin{equation*}
 Y^\alpha = \frac{\eta^\alpha}{\eta^{n+1}}
\end{equation*}
the above equation for the light cone can be written in the form
\begin{equation*}
 \xi^0 \eta^{n+1} - \eta^0 \xi^{n+1} + \Omega_{\alpha\beta} \xi^\alpha \eta^\beta =0
\end{equation*}
 which is manifestly invariant under $Sp(2n+2,\mathbb{R})$.

\section{Minimal unitary realizations of the quasiconformal groups
$\mathrm{SO}(p+2,q+2)$}

In our earlier work \cite{GP2} we constructed  the minimal unitary
representations of $SO\left(d+2, 4\right)$ obtained by quantization
of their realizations as quasiconformal groups. That construction
carries over in a straightforward manner to the other real forms
$SO\left(p+2,q+2\right)$ which we give in this section. They were
studied also in \cite{palmkvist} using the quasiconformal approach
and  in \cite{bizi} by other methods.

Now the relevant subgroup for the minimal unitary realization is
\begin{equation}
  SO(p,q)\times SO(2,2) \subset SO(p+2,q+2)
\end{equation}
where
\begin{equation}
  SO(2,2)=Sl(2,\mathbb{R}) \times Sl(2,\mathbb{R})= Sp(2,\mathbb{R})
  \times Sp(2,\mathbb{R})
\end{equation}
and one of factors above  can be identified with the distinguished
$Sl\left(2,\mathbb{R}\right)$ subgroup. The relevant 5-grading of
the Lie algebra of $SO(p+2,q+2)$ is then given as
\begin{equation}
 \mathfrak{so}(p+2,q+2) = \mathfrak{g}^{-2}  \oplus  \mathfrak{g}^{-1}  \oplus
 \left(  \mathfrak{so}(p,q)\oplus \mathfrak{sp}\left(2,
\mathbb{R}\right)  \oplus \Delta \right)
  \oplus \mathfrak{g}^{+1}  \oplus  \mathfrak{g}^{+2}
\end{equation}
where grade $\pm 1$ subspaces transform in the $[(p+q),2]$
dimensional representation of $SO(p,q)\times Sl(2,\mathbb{R})$.

Let $X^\mu$ and $P_\mu$ be canonical coordinates and momenta in
$\mathbb{R}^{(p,q)}$:
\begin{equation}
    \left[ X^\mu, P_\nu\right] = i {\delta^\mu_\nu}
\end{equation}
Also let $x$ be an additional ``cocycle''
coordinate and $p$ be its conjugate momentum:
\begin{equation}
   \left[ x, p \right] = i
\end{equation}
They are taken to satisfy the following Hermiticity conditions:
\begin{equation}
   \left(X^\mu\right)^\dagger = \eta_{\mu\nu} X^\nu \qquad
   \left(P_\mu\right)^\dagger = \eta^{\mu\nu} P_\nu \qquad
   p^\dagger = p \qquad
   x^\dagger = x
\end{equation}
where $\eta_{\mu\nu}$ is the $SO(p,q)$ invariant metric. The
subgroup $H$ of $SO(p+2,q+2)$ is $SO(p,q)\times Sp(2,\mathbb{R})_J$
whose generators we will denote as ($M_{\mu\nu}$, $J_{\pm}$, $J_0$).
The grade $-1$ generators will be denoted as ($U_{\mu}$, $V^{\mu}$)
and the grade $-2$ generator as $K_-$. The generators of $H$, its
4-th order invariant $\mathcal{I}_4$ and the negative grade
generators are realized as follows:
\begin{equation}
\begin{split}
  & \begin{aligned}
      M_{\mu\nu} &= i \eta_{\mu\rho} X^\rho P_\nu - i \eta_{\nu\rho} X^\rho P_\mu  \\
      U_\mu &= x P_\mu  \qquad V^\mu = x X^\mu \\
      K_- & = \frac{1}{2} x^2
   \end{aligned}
\qquad\qquad
    \begin{aligned}
        J_0 &= \frac{1}{2} \left( X^\mu P_\mu + P_\mu X^\mu \right) \\
        J_- &= X^\mu X^\nu \eta_{\mu\nu} \\
        J_+ &= P_\mu P_\nu \eta^{\mu\nu}
    \end{aligned} \\
   &\phantom{aga} \mathcal{I}_4 = \left(X^\mu X^\nu \eta_{\mu\nu}\right) \left(P_\mu P_\nu \eta^{\mu\nu}\right) +
    \left(P_\mu P_\nu \eta^{\mu\nu}\right) \left(X^\mu X^\nu \eta_{\mu\nu}\right) \\
     & \phantom{aga also} -  \left(X^\mu P_\mu\right)\left( P_\nu X^\nu \right) -
       \left(P_\mu X^\mu\right) \left( X^\nu P_\nu\right)
\end{split}
\end{equation}
where $\eta_{\mu\nu}$ is the flat metric with signature $(p,q)$.

It is easy to verify that the generators $M_{\mu\nu}$ and
$J_{0,\pm}$ satisfy the commutation relations of
$\mathfrak{so}\left(p,q\right)\oplus \mathfrak{sp}\left(2,
\mathbb{R}\right)$
\begin{equation}
\begin{split}
       \left[ M_{\mu\nu}, M_{\rho\tau} \right] &= \eta_{\nu\rho} M_{\mu\tau} - \eta_{\mu\rho} M_{\nu\tau} + \eta_{\mu\tau} M_{\nu\rho} - \eta_{\nu\tau} M_{\mu\rho}   \\
     \left[ J_0, J_\pm \right] &= \pm 2 i J_\pm \qquad
          \left[ J_-, J_+ \right] = 4 i J_0
\end{split}
\end{equation}
under which coordinates $X^\mu$ and momenta $P_\mu$ transform as
$SO\left(p,q\right)$ vectors and form doublets of the symplectic
group $Sp(2,\mathbb{R})_J$:
\begin{equation}
     \begin{aligned}
          \left[ J_0, V^\mu \right] &=  - i V^\mu \\
          \left[ J_0, U_\mu \right] &=  + i U_\mu
     \end{aligned}
         \quad
     \begin{aligned}
         \left[ J_-, V^\mu \right] &= 0 \\
         \left[ J_-, U_\mu \right] &= 2 i \eta_{\mu\nu} V^\nu
     \end{aligned}
          \quad
     \begin{aligned}
         \left[ J_+, V^\mu \right] &= -2 i \eta^{\mu\nu} U_\nu \\
         \left[ J_+, U_\mu \right] &= 0
     \end{aligned}
\end{equation}
The generators in the subspace $\mathfrak{g}^{-1}\oplus
\mathfrak{g}^{-2}$ form a Heisenberg algebra
\begin{equation}
    \left[ V^\mu, U_\nu \right] = 2 i {\delta^\mu}_\nu \,  K_- \,.
\end{equation}
with $K_-$ playing the role of central charge.

Using the quartic invariant we define the grade +2 generator as
\begin{equation}
    K_+ = \frac{1}{2} p^2 + \frac{1}{4 \, y^2} \left( \mathcal{I}_4 +
    \frac{(p+q-2)^2+3}{2} \right)
\end{equation}
Then the  grade $+1$ generators are obtained by commutation
relations
\begin{equation}
   \Tilde{V}^\mu = -i \left[ V^\mu, K_+ \right]  \qquad
   \Tilde{U}_\mu = -i \left[ U_\mu, K_+ \right]
\end{equation}
which explicitly read as follows
\begin{equation}
\begin{split}
    \Tilde{V}^\mu &= p X^\mu + \frac{1}{2} x^{-1}  \left(  P_\nu X^\lambda X^\rho  +
     X^\lambda X^\rho  P_\nu \right) \eta^{\mu\nu} \eta_{\lambda\rho} \\
       &- \frac{1}{4} x^{-1} \left( X^\mu \left(X^\nu P_\nu + P_\nu X^\nu\right) +
      \left(X^\nu P_\nu + P_\nu X^\nu\right) X^\mu\right) \\
    \Tilde{U}_\mu &= p P_\mu - \frac{1}{2} x^{-1}
   \left(  X^\nu P_\lambda P_\rho  +  P_\lambda P_\rho  X^\nu \right) \eta_{\mu\nu} \eta^{\lambda\rho} \\
       &+ \frac{1}{4} x^{-1} \left( P_\mu \left(X^\nu P_\nu + P_\nu X^\nu\right) +
      \left(X^\nu P_\nu + P_\nu X^\nu\right) P_\mu\right).
\end{split}
\end{equation}
Then one finds that the generators in $\mathfrak{g}^{+1} \oplus
\mathfrak{g}^{+2} $ subspace form an  Heisenberg algebra as well
\begin{equation}
  \left[ \Tilde{V}^\mu, \Tilde{U}_\nu \right] = 2 i {\delta^\mu}_\nu K_+ \qquad
 V^\mu = i \left[ \Tilde{V}^\mu, K_- \right] \qquad  U_\mu = i \left[ \Tilde{U}_\mu, K_- \right] .
\end{equation}
Commutators $\left[ \mathfrak{g}^{-1}, \mathfrak{g}^{+1} \right]$
close into $\mathfrak{g}^0$ as follows
\begin{equation}
\begin{split}
   \left[ U_\mu, \Tilde{U}_\nu \right] &= i \eta_{\mu\nu} J_- \qquad
   \left[ V^\mu, \Tilde{V}^\nu \right] = i \eta^{\mu\nu} J_+ \\
   \left[ V^\mu, \Tilde{U}_\nu \right] &= 2 \eta^{\mu\rho} M_{\rho\nu} + i {\delta^\mu}_\nu \left( J_0 +  \Delta \right)
   \\
    \left[ U_\mu, \Tilde{V}^\nu \right] &= - 2 \eta^{\nu\rho} M_{\mu\rho} + i {\delta^\nu}_\mu \left( J_0 -  \Delta \right)
\end{split}
\end{equation}
where $\Delta$ is the  generator that determines the 5-grading
\begin{equation}
   \Delta = \frac{1}{2} \left( x p + p x\right)
\end{equation}
such that
\begin{equation}
   \left[ K_-, K_+ \right] = i \Delta \qquad \left[ \Delta, K_\pm \right] = \pm 2 i K_\pm \quad
\end{equation}
\begin{equation}
 \left[ \Delta, U_\mu \right] = - i U_\mu \quad \left[ \Delta, V^\mu \right] = - i V^\mu  \quad
 \left[ \Delta, \Tilde{U}_\mu \right] =  i \Tilde{U}_\mu \quad
 \left[ \Delta, \Tilde{V}^\mu \right] =  i \Tilde{V}^\mu  \quad
\end{equation}
The quadratic Casimir operators of subalgebras
$\mathfrak{so}\left(p,q\right)$, $\mathfrak{sp}\left(2,
\mathbb{R}\right)_J$ of grade zero subspace and
$\mathfrak{sp}\left(2, \mathbb{R}\right)_K$ generated by $K_{\pm}$
and $\Delta$ are
\begin{equation}
 \begin{aligned}
   M_{\mu\nu} M^{\mu\nu} &= - \mathcal{I}_4 - 2 \left(p+q\right) \cr
   J_- J_+ +  J_+ J_- - 2  \left(J_0\right)^2 &= \mathcal{I}_4 + \frac{1}{2}\left(p+q \right)^2 \cr
    K_- K_+ + K_+ K_- - \frac{1}{2} \Delta^2 &= \frac{1}{4} \mathcal{I}_4 + \frac{1}{8} \left(p+q\right)^2
 \end{aligned}
\end{equation}
Note that they all reduce to $\mathcal{I}_4$ modulo some additive
and multiplicative  constants. Noting also that
\begin{equation}
  \left( U_\mu \Tilde{V}^\mu + \Tilde{V}^\mu U_\mu  - V^\mu \Tilde{U}_\mu - \Tilde{U}_\mu V^\mu \right) =
     2 \mathcal{I}_4 + \left(p+q\right)\left(p+q+4\right)
\end{equation}
we conclude that there exists a family of degree 2 polynomials in
the enveloping algebra of $\mathfrak{so}\left(p+2,q+2\right)$ that
degenerate to a c-number for the  minimal unitary  realization, in
accordance with Joseph's theorem \cite{Joseph}:
\begin{equation}\label{eq:JosephIdeal1}
\begin{split}
   M_{\mu\nu} M^{\mu\nu} &+ \kappa_1 \left( J_- J_+ +  J_+ J_- - 2  \left(J_0\right)^2 \right)
   +  4 \kappa_2 \left(K_- K_+ + K_+ K_- - \frac{1}{2} \Delta^2\right) \\&-
     \frac{1}{2}\left(\kappa_1+\kappa_2-1\right)
    \left( U_\mu \Tilde{V}^\mu + \Tilde{V}^\mu U_\mu  - V^\mu \Tilde{U}_\mu - \Tilde{U}_\mu V^\mu \right)
  \\ &= \frac{1}{2}\left(p+q\right)\left( p+q - 4\left(\kappa_1+\kappa_2\right)
  \right)
\end{split}
\end{equation}
The quadratic Casimir of $\mathfrak{so}\left(p+2, q+2\right)$
corresponds to the choice $2 \kappa_1 = 2 \kappa_2 = -1$ in
\eqref{eq:JosephIdeal1}. Hence the eigenvalue of the quadratic
Casimir for the minimal unitary representation is equal to
$\frac{1}{2} \left(p+q\right) \left(p+q+4\right)$.

\section{Minimal unitary realizations of the quasiconformal
groups $\mathrm{SO}^\ast(2n+4)$}

The noncompact group $SO^\ast(2n+4)$ is a subgroup of
$SL(2n+4,\mathbb{C})$ whose maximal compact subgroup is $U(n+2)$. We
have the inclusions
\begin{equation}
 SO^\ast(2n+4) \subset SU^\ast(2n+4) \subset SL(2n+4,\mathbb{C})
\end{equation}
As a matrix group $SU^\ast(2n+4)$ is generated by matrices $U$
belonging to $SL(2n+4,\mathbb{C})$ that satisfy
\begin{equation}
  U \mathbb{J}=\mathbb{J}U^\ast
\end{equation}
where $\mathbb{J}$ is a $(2n+4)\times (2n+4)$ matrix that is
antisymmetric
\begin{equation}
 \mathbb{J}^T=-\mathbb{J}
\end{equation}
and whose square is the identity matrix
\begin{equation}
   \mathbb{J}^2=-\mathbb{I}
\end{equation}
The matrices $U$ belonging to the subgroup $SO^*(2n+4)$ of
$SU^*(2n+4)$ satisfy, in addition, the condition
\begin{equation}
 U U^T=\mathbb{I}
\end{equation}
Thus $SO^\ast(2n+4)$ leaves invariant both the Euclidean metric
$\delta_{IJ}$ and the complex structure
$\mathbb{J}_{IJ}=-\mathbb{J}_{JI}$ where $I,J,..=1,2,...2n+4$. Hence
$SO^\ast(2n+4)$ is also a subgroup of the complex rotation group
$SO(2n+4,\mathbb{C})$.

To obtain the 5-grading of the Lie algebra of $SO^\ast(2n+4)$ so as to
construct its minimal unitary representation we need to consider its
decomposition with respect to its subgroup
\begin{equation}
  SO^\ast(2n)\times SO^\ast(4) \subset SO^\ast(2n+4)
\end{equation}
where
\begin{equation}
  SO^\ast(4)= SU(2)\times SL(2,\mathbb{R})
\end{equation}
The distinguished $SL(2,\mathbb{R})$ subgroup can then be identified
with the factor $SL(2,\mathbb{R})$ above. The corresponding
5-grading of the Lie algebra of $SO^\ast(2n+4)$ is then
\begin{equation}
 \mathfrak{so}^\ast(2n+4) = \mathfrak{g}^{-2}  \oplus  \mathfrak{g}^{-1}  \oplus
 \left(  \mathfrak{so}^\ast(2n) \oplus \mathfrak{su}(2)  \oplus \Delta \right)
  \oplus \mathfrak{g}^{+1}  \oplus  \mathfrak{g}^{+2}
\end{equation}
where grade $\pm 1$ subspaces transform in the $[2n,2]$ dimensional
representation of $SO^\ast(2n)\times SU(2)$.
 Let $X^\mu$ and $P_\mu$ be canonical coordinates and momenta in
$\mathbb{R}^{(2n)}$:
\begin{equation}
    \left[ X^\mu, P_\nu\right] = i {\delta^\mu_\nu}
\end{equation}
They satisfy the Hermiticity conditions
\begin{equation}
  \left( X^{\mu} \right )^{\dagger}= \mathbb{J}_{\mu\nu} X^{\nu}
\end{equation}
\begin{equation}
 \left( P_{\mu} \right )^{\dagger}=   \mathbb{J}^{\mu\nu} P_{\nu}
\end{equation}
Let $x$ be an additional ``cocycle'' coordinate and $p$ be its conjugate momentum:
\begin{equation}
   \left[ x, p \right] = i
\end{equation}
as in the previous section. The subgroup $H$ is now $SO^*(2n)\times
SU(2)_J$ whose generators are ($M_{\mu\nu}$, $J_{\pm}$, $J_0$). The
grade $-1$ generators will be denoted as ($U_{\mu}$, $V^{\mu}$) and
the grade $-2$ generator as $K_-$  as in the previous section. The
generators of $H$, its 4-th order invariant $\mathcal{I}_4$ and the
negative grade generators are realized as follows:
\begin{equation}
\begin{split}
  & \begin{aligned}
      M_{\mu\nu} &= i \delta_{\mu\rho} X^\rho P_\nu - i \delta_{\nu\rho} X^\rho P_\mu  \\
      U_\mu &= x P_\mu  \qquad V^\mu = x X^\mu \\
      K_- & = \frac{1}{2} x^2
   \end{aligned}
\qquad\qquad
    \begin{aligned}
        J_0 &= \frac{1}{2} \left( X^\mu P_\mu + P_\mu X^\mu \right) \\
        J_- &= X^\mu X^\mu  \\
        J_+ &= P_\mu P_\mu
    \end{aligned} \\
   &\phantom{aga} \mathcal{I}_4 = \left(X^\mu X^\mu \right) \left(P_\nu P_\nu \right) +
    \left(P_\mu P_\mu \right) \left(X^\nu X^\nu \right) \\
     & \phantom{aga also} -  \left(X^\mu P_\mu\right)\left( P_\nu X^\nu \right) -
       \left(P_\mu X^\mu\right) \left( X^\nu P_\nu\right)
\end{split}
\end{equation}
where $\delta_{\mu\nu}$ is the flat Euclidean metric in $2n$
dimensions.

It is easy to verify that the generators $M_{\mu\nu}$ and
$J_{0,\pm}$ satisfy the commutation relations of
$\mathfrak{so}^\ast\left(2n\right)\oplus \mathfrak{su}\left(2\right)_J$:
\begin{equation}
\begin{split}
       \left[ M_{\mu\nu}, M_{\rho\tau} \right] &= \delta_{\nu\rho} M_{\mu\tau} - \delta_{\mu\rho} M_{\nu\tau} + \eta_{\mu\tau} M_{\nu\rho} - \eta_{\nu\tau} M_{\mu\rho}   \\
     \left[ J_0, J_\pm \right] &= \pm 2 i J_\pm \qquad
          \left[ J_-, J_+ \right] = 4 i J_0
\end{split}
\end{equation}
under which coordinates $X^\mu$ ($V^{\mu}$) and momenta $P^\mu$
($U^{\mu}$)  transform as vectors of $SO^*(2n)$ and form doublets of
$SU(2)$:
\begin{equation}
     \begin{aligned}
          \left[ J_0, V^\mu \right] &=  - i V^\mu \\
          \left[ J_0, U_\mu \right] &=  + i U_\mu
     \end{aligned}
         \quad
     \begin{aligned}
         \left[ J_-, V^\mu \right] &= 0 \\
         \left[ J_-, U_\mu \right] &= 2 i  V_\mu
     \end{aligned}
          \quad
     \begin{aligned}
         \left[ J_+, V^\mu \right] &= -2 i  U^\nu \\
         \left[ J_+, U_\mu \right] &= 0
     \end{aligned}
\end{equation}
The generators in the subspace $\mathfrak{g}^{-1}\oplus
\mathfrak{g}^{-2}$ form a Heisenberg algebra
\begin{equation}
    \left[ V^\mu, U_\nu \right] = 2 i {\delta^\mu}_\nu \,  K_- \,.
\end{equation}
with $K_-$ playing the role of ``$\hbar$''.

Using the quartic invariant we define the grade +2 generator as
\begin{equation}
    K_+ = \frac{1}{2} p^2 + \frac{1}{4 \, y^2} \left( \mathcal{I}_4 +
    \frac{4 (n-1)^2+3}{2} \right)
\end{equation}
Then the  grade $+1$ generators are obtained by commutation
relations
\begin{equation}
   \Tilde{V}^\mu = -i \left[ V^\mu, K_+ \right]  \qquad
   \Tilde{U}_\mu = -i \left[ U_\mu, K_+ \right]
\end{equation}
which explicitly read as follows
\begin{equation}
\begin{split}
    \Tilde{V}^\mu &= p X^\mu + \frac{1}{2} x^{-1}  \left(  P_\nu X^\lambda X^\rho  +
     X^\lambda X^\rho  P_\nu \right) \eta^{\mu\nu} \eta_{\lambda\rho} \\
       &- \frac{1}{4} x^{-1} \left( X^\mu \left(X^\nu P_\nu + P_\nu X^\nu\right) +
      \left(X^\nu P_\nu + P_\nu X^\nu\right) X^\mu\right) \\
    \Tilde{U}_\mu &= p P_\mu - \frac{1}{2} x^{-1}
   \left(  X^\nu P_\lambda P_\rho  +  P_\lambda P_\rho  X^\nu \right) \eta_{\mu\nu} \eta^{\lambda\rho} \\
       &+ \frac{1}{4} x^{-1} \left( P_\mu \left(X^\nu P_\nu + P_\nu X^\nu\right) +
      \left(X^\nu P_\nu + P_\nu X^\nu\right) P_\mu\right) .
\end{split}
\end{equation}
Then one finds that the generators in $\mathfrak{g}^{+1} \oplus
\mathfrak{g}^{+2} $ subspace form an  Heisenberg algebra as well
\begin{equation}
  \left[ \Tilde{V}^\mu, \Tilde{U}_\nu \right] = 2 i {\delta^\mu}_\nu K_+ \qquad
 V^\mu = i \left[ \Tilde{V}^\mu, K_- \right] \qquad  U_\mu = i \left[ \Tilde{U}_\mu, K_- \right] .
\end{equation}
Commutators $\left[ \mathfrak{g}^{-1}, \mathfrak{g}^{+1} \right]$
close into $\mathfrak{g}^0$ as follows
\begin{equation}
\begin{split}
   \left[ U_\mu, \Tilde{U}_\nu \right] &= i \eta_{\mu\nu} J_- \qquad
   \left[ V^\mu, \Tilde{V}^\nu \right] = i \eta^{\mu\nu} J_+ \\
   \left[ V^\mu, \Tilde{U}_\nu \right] &= 2 \eta^{\mu\rho} M_{\rho\nu} + i {\delta^\mu}_\nu \left( J_0 +  \Delta \right)
   \\
    \left[ U_\mu, \Tilde{V}^\nu \right] &= - 2 \eta^{\nu\rho} M_{\mu\rho} + i {\delta^\nu}_\mu \left( J_0 -  \Delta \right)
\end{split}
\end{equation}
where $\Delta$ is the  generator that determines the 5-grading
\begin{equation}
   \Delta = \frac{1}{2} \left( x p + p x\right)
\end{equation}
such that
\begin{equation}
   \left[ K_-, K_+ \right] = i \Delta \qquad \left[ \Delta, K_\pm \right] = \pm 2 i K_\pm \quad
\end{equation}
\begin{equation}
 \left[ \Delta, U_\mu \right] = - i U_\mu \quad \left[ \Delta, V^\mu \right] = - i V^\mu  \quad
 \left[ \Delta, \Tilde{U}_\mu \right] =  i \Tilde{U}_\mu \quad
 \left[ \Delta, \Tilde{V}^\mu \right] =  i \Tilde{V}^\mu  \quad
\end{equation}
The quadratic Casimir operators of subalgebras
$\mathfrak{so}^\ast\left(2 n\right)$, $\mathfrak{su}\left(2\right)_J$ of grade zero subspace and
$\mathfrak{sp}\left(2, \mathbb{R}\right)_K$ generated by $K_{\pm}$
and $\Delta$ are
\begin{equation}
 \begin{aligned}
   M_{\mu\nu} M^{\mu\nu} &= - \mathcal{I}_4 - 2 \left(p+q\right) \cr
   J_- J_+ +  J_+ J_- - 2  \left(J_0\right)^2 &= \mathcal{I}_4 + \frac{1}{2}\left(p+q \right)^2 \cr
    K_- K_+ + K_+ K_- - \frac{1}{2} \Delta^2 &= \frac{1}{4} \mathcal{I}_4 + \frac{1}{8} \left(p+q\right)^2
 \end{aligned}
\end{equation}
Note that they all reduce to $\mathcal{I}_4$ modulo some additive
and multiplicative  constants. Noting also that
\begin{equation}
  \left( U_\mu \Tilde{V}^\mu + \Tilde{V}^\mu U_\mu  - V^\mu \Tilde{U}_\mu - \Tilde{U}_\mu V^\mu \right) =
     2 \mathcal{I}_4 + \left(p+q\right)\left(p+q+4\right)
\end{equation}
we conclude that there exists a family of degree 2 polynomials in
the enveloping algebra of $\mathfrak{so}^\ast\left(2n+4\right)$ that
degenerate to a c-number for the  minimal unitary  realization, in
accordance with Joseph's theorem \cite{Joseph}:
\begin{equation}\label{eq:JosephIdeal}
\begin{split}
   M_{\mu\nu} M^{\mu\nu} &+ \kappa_1 \left( J_- J_+ +  J_+ J_- - 2  \left(J_0\right)^2 \right)
   +  4 \kappa_2 \left(K_- K_+ + K_+ K_- - \frac{1}{2} \Delta^2\right) \\&-
     \frac{1}{2}\left(\kappa_1+\kappa_2-1\right)
    \left( U_\mu \Tilde{V}^\mu + \Tilde{V}^\mu U_\mu  - V^\mu \Tilde{U}_\mu - \Tilde{U}_\mu V^\mu \right)
  \\ &= n \left( 2 n - 4\left(\kappa_1+\kappa_2 -1 \right) \right)
\end{split}
\end{equation}

The quadratic Casimir of $\mathfrak{so}^\ast\left(2 n +4\right)$
corresponds to the choice $2 \kappa_1 = 2 \kappa_2 = -1$ in
\eqref{eq:JosephIdeal}. Hence the eigenvalue of the quadratic
Casimir for the minimal unitary representation is equal to
$ 2 n \left(n+2\right)$.

\section{Minimal unitary realizations of the quasiconformal
groups $SU\left(n+1, m+1\right)$} The Lie algebra
$\mathfrak{su}\left(n+1, m+1\right)$ admits the following five
graded decomposition with respect to its subalgebra
$\mathfrak{su}\left(n,m\right)$:
\begin{equation*}
   \mathfrak{su}\left(n+1, m+1\right) = 1 \oplus 2(n+m) \oplus \left( \mathfrak{su}\left(n, m\right) \oplus
        \mathfrak{u}\left(1\right) \right) \oplus 2(n+m) \oplus 1
\end{equation*}
It is realized by means of $m+n$ pairs of creation and annihilation
operators subject to the following Hermiticity condition:
\begin{equation}
   \left( a^p \right)^\dagger = \eta^{pq} a_q  \qquad
    \left[ a_q, a^p \right] = {\delta^p}_q\,.
\end{equation}

Following the steps laid down in previous sections we define
generators of $H$ as bilinears in creation and annihilation
operators
\begin{subequations}
\begin{equation}
   {J^p}_q = a^p a_q - \frac{1}{m+n} {\delta^p}_q a^r a_r
\end{equation}
Negative grade generators are
\begin{equation}
   E = \frac{1}{2} x^2 \quad E^p = x a^p \quad E_q = x a_q
\end{equation}
The quartic invariant $\mathcal{I}_4$ is related to the quadratic
Casimir of $H$ simply
\begin{equation}
  \mathcal{I}_4 = \frac{2 \left(m+n\right)}{m+n-1}\, {J^p}_q {J^q}_p + \frac{1}{2} \left( \left(m+n\right)^2 - 1\right)
\end{equation}
where the additive constant was determined such that
\begin{equation}
   F = \frac{1}{2} p^2 + \frac{1}{4} \frac{1}{x^2} \mathcal{I}_4
\end{equation}
The positive grade $\mathfrak{g}^{+1}$ generators are then found by
commuting $F$ with generators of $\mathfrak{g}^{-1}$
\begin{equation}
   F^p = - i \left[ E^p, F \right] \quad F_q = - i \left[ E_q, F \right]
\end{equation}
The $\mathfrak{u}(1)$ generator of grade 0 subalgebra is also
bilinear in oscillators
\begin{equation}
   U = \frac{1}{2} \left( a^p a_p + a_p a^p \right)
\end{equation}
\end{subequations}
Quadratic Casimir of the algebra in this realization reduces to a
c-number
\begin{equation}
\begin{split}
  \mathcal{C}_2 =&  -\frac{1}{6} {J^p}_q {J^q}_p + \frac{1}{12} \Delta^2 - \frac{1}{6}\left( E F + F E\right) - \frac{1}{12} \frac{m+n+2}{m+n} U^2 \\
   & - \frac{i}{12} \left( E_p F^p + F^p E_p - F_p E^p - E^p F_p \right) \\
   =& \,\frac{1}{24} \left(m+n+2\right)\left(m+n+1\right)
\end{split}
\end{equation}
 Positive and negative grades generators transform in the
$(n+m)^{+1} \oplus (\overline{n+m})^{-1} \oplus 1^{0}$
representation of $H$ and satisfy
\begin{equation}
\begin{split}
 \left[ {J^p}_q, {J^s}_t \right] &= {\delta^s}_q {J^p}_t - {\delta^p}_t {J^s}_q \\
 \left[ {J^p}_q, E^s \right] &= {\delta^s}_q E^p - \frac{1}{n+m} {\delta^p}_q E^s \\
 \left[ {J^p}_q, F^s \right] &= {\delta^s}_q F^p - \frac{1}{n+m} {\delta^p}_q F^s \\
 \left[ {J^p}_q, E_s \right] &= -{\delta^p}_s E_q + \frac{1}{n+m} {\delta^p}_q E_s \\
 \left[ {J^p}_q, F_s \right] &= -{\delta^p}_s F_q + \frac{1}{n+m} {\delta^p}_q F_s
\end{split}
\end{equation}
\begin{equation}
  \left[ U, E^p \right] = E^p \quad \left[ U, F^p \right] = F^p \quad \left[ U, E_p \right] = -E_p \quad
   \left[ U, F_p \right] = -F_p
\end{equation}
\begin{equation}
  \left[ \Delta, E^p \right] = -i E^p \quad \left[ \Delta, E_p \right] = -i E_p \quad
   \left[ \Delta, F^p \right] = i F^p \quad
   \left[ \Delta, F_p \right] = i F_p
\end{equation}
\begin{equation}
   \left[ \Delta, F\right] = 2 i F \qquad \left[ \Delta, E\right] = -2 i E \qquad \left[ E, F\right] = i \Delta
\end{equation}
The remaining non-zero commutation relations are as follows:
\begin{equation}
   \left[ E_p, E^q\right] = 2 E  \qquad \left[ F_p, F^q\right] = 2 F
\end{equation}
\begin{equation}
\begin{split}
  \left[ E^p, F_q \right] &= 2 i {J^p}_q + \frac{m+n+2}{m+n} i {\delta^p}_q U - {\delta^p}_q \Delta \\
  \left[ E_p, F^q \right] &= -2 i {J^q}_p - \frac{m+n+2}{m+n} i {\delta^q}_p U - {\delta^p}_q \Delta
\end{split}
\end{equation}
\begin{equation}
  \left[ F, E_p \right] = - i F_p \quad \left[ F, E^p \right] = - i F^p \quad
  \left[ E, F_p \right] =  i E_p \quad \left[ E, F^p \right] = i E^p \quad
\end{equation}

\section{Minimal unitary realizations of the quasiconformal
groups $SL\left(n+2, \mathbb{R}\right)$}

The construction of the minimal unitary realization of the
quasiconformal algebra $\mathfrak{sl}(n+2, \mathbb{R})$ traces the
same steps as in the previous section. The five-graded decomposition
is as follows
\begin{equation*}
   \mathfrak{sl}\left(n+2, \mathbb{R}\right) = 1 \oplus \left(n \oplus \Tilde{n}\right) \oplus
       \left( \mathfrak{gl}\left(n, \mathbb{R}\right) \oplus \Delta \right) \oplus
      \left(n \oplus \Tilde{n}\right) \oplus 1
\end{equation*}
Since $n \oplus \Tilde{n}$ is a direct sum of two inequivalent
self-conjugate vector representations of
$\mathfrak{gl}\left(n,\mathbb{R} \right)$, we use coordinates
$X^\mu$ and momenta $P_\mu$ as oscillator generators, where $\mu =
1, \ldots, n$, with canonical commutation relations:
\begin{equation}
   \left[ X^\mu, P_\nu \right] = i {\delta^\mu}_\nu
\end{equation}
Generators of $\mathfrak{gl}\left(n,\mathbb{R} \right)$
\begin{equation}
    {\mathcal{L}^\mu}_\nu =\frac{i}{2} \left( X^\mu P_\nu  + P_\nu X^\mu \right)
\end{equation}
have the following commutation relations
\begin{equation}
   \left[ {\mathcal{L}^\mu}_\nu, {\mathcal{L}^\tau}_\rho \right] = {\delta^\tau}_\nu {\mathcal{L}^\mu}_\rho -
         {\delta^\mu}_\rho {\mathcal{L}^\tau}_\nu
\end{equation}
The one-dimensional center of the reductive algebra
$\mathfrak{gl}\left(n,\mathbb{R} \right)$ is spanned by
\begin{equation}
    U = \sum_{\mu=1}^n {\mathcal{L}^\mu}_\mu
\end{equation}
The quadratic Casimir of $\mathfrak{gl}\left(n,\mathbb{R} \right)$
is given simply as a trace of $\mathcal{L}^2$:
\begin{equation}
  \mathcal{C}_2\left( \mathfrak{gl}\left(n,\mathbb{R} \right) \right) = {\mathcal{L}^\mu}_\nu  {\mathcal{L}^\nu}_\mu = \frac{n}{4} - \left(X^\mu P_\mu\right)\left(P_\nu X^\nu\right)
\end{equation}
Generators of the negative grades
\begin{equation}
 E = \frac{1}{2} {x^2} \qquad E^\mu = x X^\mu \qquad E_\nu = x P_\nu
\end{equation}
form the Heisenberg algebra
\begin{equation}
\begin{split}
    &\left[ E^\mu, E_\nu \right] = \left(2 i E\right) {\delta^\mu}_\nu  \qquad
    \left[ E, E^\mu\right] = 0 \\
    & \left[ E^\nu, E^\mu\right] = 0 \qquad  \left[ E, E_\mu \right] = 0 \qquad \left[ E_\nu, E_\mu\right] = 0
\end{split}
\end{equation}
The generator of grade +2 subspace takes on the familiar form
\begin{subequations}
\begin{equation}
  F = \frac{1}{2} p^2 + \frac{1}{2} \frac{1}{x^2} \mathcal{I}_4 = \frac{1}{2} p^2 + \frac{1}{2} \frac{1}{x^2} \left(
   \frac{n^2-1}{4} - \left(X^\mu P_\mu\right)\left(P_\nu X^\nu\right)  \right)
\end{equation}
and leads to  the following grade +1 generators
\begin{equation}
   F^\mu = - i \left[ x X^\mu, F \right] = p X^\mu - \frac{1}{2} \frac{1}{x} \left( X^\mu \left(P_\nu X^\nu\right) +
     \left(X^\nu P_\nu\right) X^\mu \right)
\end{equation}
\begin{equation}
   F_\mu = - i \left[ x P_\mu, F \right] = p P_\mu + \frac{1}{2} \frac{1}{x} \left( P_\mu \left(P_\nu X^\nu\right) +
     \left(X^\nu P_\nu\right) P_\mu \right)
\end{equation}
\end{subequations}
They form the dual Heisenberg algebra
\begin{equation}
\begin{split}
    &\left[ F^\mu, F_\nu \right] = \left(2 i F\right) {\delta^\mu}_\nu  \qquad
    \left[ F, F^\mu\right] = 0 \\
    & \left[ F^\nu, F^\mu\right] = 0 \qquad  \left[ F, F_\mu \right] = 0 \qquad \left[ F_\nu, F_\mu\right] = 0
\end{split}
\end{equation}
Subspaces $\mathfrak{g}^{\pm 1}$ transform under
$\mathfrak{gl}\left(n, \mathbb{R}\right)$ as $\mathbf{n} \oplus
\Tilde{\mathbf{n}}$ each:
\begin{equation}
\begin{split}
 &\left[ {\mathcal{L}^\mu}_\nu, E^\rho \right] = {\delta^\rho}_\nu E^\mu \qquad
 \left[ {\mathcal{L}^\mu}_\nu, E_\rho \right] = -{\delta^\mu}_\rho E_\nu \\
 &\left[ {\mathcal{L}^\mu}_\nu, F^\rho \right] = {\delta^\rho}_\nu F^\mu \qquad
 \left[ {\mathcal{L}^\mu}_\nu, F_\rho \right] = -{\delta^\mu}_\rho F_\nu
\end{split}
\end{equation}
Other cross grade commutation relations read
\begin{equation}
 \left[ E, F^\mu\right] = i E^\mu \quad \left[ E, F_\mu\right] = i E_\mu \quad
 \left[ F, E^\mu\right] = -i F^\mu \quad \left[ F, E_\mu\right] = -i F_\mu
\end{equation}
\begin{equation}
  \left[ E, F\right] = i \Delta \qquad \left[ \Delta, E \right] = - 2 i E \qquad
  \left[ \Delta, F \right] = + 2 i F \qquad
\end{equation}
\begin{equation}
\begin{split}
  \left[ \Delta, E^\mu \right]  = - i E^\mu \qquad \left[ \Delta, E_\mu \right]  = - i E_\mu \\
  \left[ \Delta, F^\mu \right]  = + i F^\mu \qquad \left[ \Delta, F_\mu \right]  = + i F_\mu
\end{split}
\end{equation}
\begin{equation}
  \left[ E^\mu, F^\nu \right] = 0 \qquad \left[ E_\mu, F_\nu \right] = 0
\end{equation}
\begin{equation}
\begin{split}
  \left[ E^\mu, F_\nu \right] &= 2 {\mathcal{L}^\mu}_\nu + {\delta^\mu}_\nu \left(  U + i \Delta \right)
 \\
   \left[ E_\mu, F^\nu \right] &= 2 {\mathcal{L}^\nu}_\mu + {\delta^\nu}_\mu \left( U - i \Delta \right)
\end{split}
\end{equation}
A short calculation verifies that the quadratic Casimir of
$\mathfrak{sl}\left(n+2, \mathbb{R}\right)$ is
\begin{equation}
\begin{split}
   C_2 &= {\mathcal{L}^\mu}_\nu {\mathcal{L}^\nu}_\mu +
   \frac{1}{2} \left({\mathcal{L}^\mu}_\mu\right)\left( {\mathcal{L}^\nu}_\nu \right) - \frac{1}{2} \Delta^2 +
   \left( E F + F E\right) \\
    & + \frac{1}{2} \left( E^\mu F_\mu + F_\mu E^\mu - F^\mu E_\mu - E_\mu F^\mu \right)
\end{split}
\end{equation}
When evaluated on the quasi-conformal realization it reduces to c-number:
\begin{equation}
    C_2 = -\frac{1}{4} \left(n+2\right)\left(n+1\right)
\end{equation}
Quadratic Casimir is just an element of the Joseph ideal as follows
from relations below
\begin{equation}
  {\mathcal{L}^\mu}_\nu {\mathcal{L}^\nu}_\mu +
   \frac{1}{2} \left({\mathcal{L}^\mu}_\mu\right)\left( {\mathcal{L}^\nu}_\nu \right) = \frac{3}{2} \mathcal{I}_4 +
   \frac{1}{8} \left(3 - 2 n\left(n-1\right) \right)
\end{equation}
\begin{equation}
- \frac{1}{2} \Delta^2 +  \left( E F + F E\right)
 = \frac{1}{2} \mathcal{I}_4 -
   \frac{3}{8}
\end{equation}
\begin{equation}
\frac{1}{2} \left( E^\mu F_\mu + F_\mu E^\mu - F^\mu E_\mu - E_\mu
F^\mu \right)
 = -2 \mathcal{I}_4 - n + \frac{1}{2}
\end{equation}
Namely, any linear combination of the above three expressions
collapses to a c-number provided coefficients are matched as to
cancel all $\mathcal{I}_4$.

\section{Minimal Unitary Realizations  of Lie Superalgebras}
In this section we will extend the construction of the minimal
unitary representations of Lie groups obtained by quantization  of
their quasi conformal realizations to the construction of the
minimal representations of Lie superalgebras. In analogy with the
Lie algebras we consider  5-graded  simple Lie superalgebras
\begin{equation}
 \mathfrak{g}^{-2}_B  \oplus  \mathfrak{g}^{-1}_F  \oplus \left(  \mathfrak{g}^{0}  \oplus \Delta \right)_B
 \oplus  \mathfrak{g}^{+1}_F  \oplus  \mathfrak{g}^{+2}_B
\end{equation}
where $\mathfrak{g}^{\pm 2}$ are 1-dimensional subspaces each, and
$\mathfrak{g}^{-2} \oplus \Delta \oplus \mathfrak{g}^{+2}$ form
$\mathfrak{sl}(2\,, \mathbb{R} )$. In this paper we restrict
ourselves  to Lie superalgebras  whose grade $\pm 1$  generators are
all  odd ( exhaustively).
 Let $J^a$ denote generators of $\mathfrak{g}^0$
\begin{equation}
   \left[ J^a \,, J^b \right] = {f^{ab}}_c J^c
\end{equation}
and let $\rho$ denote the irreducible orthogonal representation with
a definite Dynkin index by which $\mathfrak{g}^0$ acts on
$\mathfrak{g}^{\pm 1}$
\begin{equation}
   \left[ J^a \,, E^{\alpha} \right] = {\lambda^{a\alpha}}_\beta E^\beta
\qquad
   \left[ J^a \,, F^{\alpha} \right] = {\lambda^{a\alpha}}_\beta F^\beta
\end{equation}
where $E^\alpha$ are odd generators that span $\mathfrak{g}^{-1}$
\begin{equation}
   \left\{ E^\alpha \,, E^\beta  \right\} = \Omega_s^{\alpha\beta} E
\end{equation}
and $F^\alpha$ generators that span $\mathfrak{g}^{+1}$
\begin{equation}
   \left\{ F^\alpha \,, F^\beta  \right\} = \Omega_s^{\alpha\beta} F
\end{equation}
and $\Omega_s^{\alpha\beta}$ is now  a symmetric invariant tensor.
Hence negative (positive) grade subspace form a super Heisenberg
algebra. Due to 5-graded structure we can impose
\begin{equation}
  F^\alpha = \left[ E^\alpha \,, F \right] \qquad
  E^\alpha = \left[ E \,, F^\alpha \right]
\end{equation}

\noindent Now we realize the generators using anti-commuting
covariant oscillators $\xi^\alpha$
\begin{equation}
   \left\{ \xi^\alpha \,, \xi^\beta \right\} = \Omega_s^{\alpha\beta}
\end{equation}
plus an extra bosonic coordinate $y$ and its conjugate momentum $p$.
The non-positive grade generators take the form \footnote{We are
following conventions of \cite{bg97} for superalgebras  as well.}
\begin{equation}
  E = \frac{1}{2} y^2 \qquad E^\alpha = y \xi^\alpha \qquad
 J^a =  - \frac{1}{2}  {\lambda^a}_{\alpha\beta} \xi^\alpha \xi^\beta
\end{equation}
 The quadratic Casimir
of $\mathfrak{g}^0$ is taken to be
\begin{equation}
 \mathcal{C}_2\left(\mathfrak{g}^0\right) = \eta_{ab} J^a J^b
\end{equation}
and the grade +2 generator $F$ is assumed to be of the form
\begin{equation} \label{eq:exprFf}
   F = \frac{1}{2} p^2 + \kappa y^{-2} \left( \mathcal{C}_2 + \mathfrak{C} \right)
\end{equation}
for some constants $\kappa$ and $\mathfrak{C}$ to be determined
later. Commuting $E$ with $F^{\alpha}$ we obtain
\begin{equation}
  F^\alpha = i p \, \xi^\alpha + \kappa y^{-1} \left[ \xi^\alpha \,, \mathcal{C}_2 \right]
\end{equation}

By inspection we have
\begin{equation}
  \left[ \xi^\alpha \,, \mathcal{C}_2 \right] = - 2 \,
   {\left(\lambda^{a}\right)^{\alpha}}_\beta \xi^\beta J_a + \, C_\rho \, \xi^\alpha
  \label{eq:aux1f}
\end{equation}
and we shall use the following Ansatz \cite{bg97}
\begin{equation}
   {\left(\lambda^a\right)^\beta}_\gamma {\left(\lambda_a\right)^\alpha}_\delta +
   \left( \lambda^a\right)^{\beta\alpha} \left(\lambda_a\right)_{\gamma\delta} =
   \frac{C_\rho}{N-1} \left( \Omega_s^{\alpha\beta} \Omega_{s\gamma\delta} +
           {\delta^\beta}_\gamma {\delta^\alpha}_\delta -
           2 {\delta^\alpha}_\gamma {\delta^\beta}_\delta \right)
\end{equation}
to calculate the remaining super commutation relations.

For the anticommutators of grade +1 generators with grade -1
generators we get
\begin{eqnarray}
\label{eq:EFf}
   \left\{ E^\alpha \,, F^\beta \right\} &=& i \left( y \, p \right) \Omega_s^{\alpha\beta}
     + \xi^\beta \, \xi^\alpha + \kappa \left\{ \xi^\alpha \,, \left[ \xi^\beta \,,\mathcal{C}_2 \right] \right\}
\\ \nonumber
   \left\{ E^\alpha,\,F^\beta \right\}& =& -\Omega_s^{\alpha\beta} \Delta  -
   6 \kappa \left(\lambda^a\right)^{\beta\alpha} J_a +
   \left( \frac{3 \kappa C_\rho}{N - 1} - \frac{1}{2} \right) \left( \xi^\beta \xi^\alpha -
   \xi^\alpha \xi^\beta \right).
\end{eqnarray}
Now the bilinears $ \left( \xi^\beta \xi^\alpha -
   \xi^\alpha \xi^\beta \right) $ on the right hand side generate the Lie algebra $\mathfrak{so}(N)$.
Therefore, for those Lie superalgebras whose grade zero subalgebras
$g^0$ are different from $\mathfrak{so}(N)$ we must impose the
constraint:
\begin{equation}\label{eq:firstsuper}
\left( \frac{3 \kappa C_\rho}{N - 1} - \frac{1}{2} \right)
\end{equation}
 For the anticommutators $\left\{ F^\alpha,\, F^\beta
\right\}$ we get
\begin{subequations}
\label{eq:FFf}
\begin{multline}
   \left\{ F^\alpha \,, F^\beta \right\} = - p^2 \Omega_s^{\alpha\beta} \\-
   \frac{\kappa}{y^{2}} \left(   \xi^\alpha \left[ \xi^\beta \,, \mathcal{C}_2 \right] +
          \xi^\beta \left[ \xi^\alpha \,, \mathcal{C}_2 \right] -
          \kappa \left\{ \left[ \xi^\alpha \,, \mathcal{C}_2 \right]  ,
                        \left[ \xi^\beta \,, \mathcal{C}_2 \right]
                 \right\}
   \right)
\end{multline}
\begin{equation}
\begin{split}
   \left\{ F^\alpha,\,F^\beta\right\} & = - 2 F \Omega_s^{\alpha\beta} =
      - 2 \left( \frac{p^2}{2} + \frac{k}{x^2} \left( \frac{1}{2}\kappa C_\rho^2 + \frac{1}{2} C_\rho \right)
         \right)\Omega_s^{\alpha\beta}
   \\  &- \frac{\kappa}{x^2} \left(  - 4 \left( \xi^\alpha {\left( \lambda_a\right)^\beta}_\gamma +
                                                 \xi^\beta {\left( \lambda_a\right)^\alpha}_\gamma
                                         \right) \xi^\gamma J^a +
        12 \kappa \left( \lambda_a \lambda_b \right)^{\alpha\beta} J^b J^a  \right.  \\ &\left.
       + 8 \kappa \left( \lambda_a \lambda_b \right)^{\beta\alpha} J^b J^a
        - 4 \kappa {\left( \lambda_a \right)^\alpha}_\delta {\left( \lambda_b \right)^\beta}_\gamma
             \xi^\delta \xi^\gamma {f^{ab}}_c J^c
                             \right)
\end{split}
\end{equation}
\end{subequations}
Taking the $\Omega_s$ trace we obtain
\begin{equation}
   N = 8 - 10 \kappa i_\rho \ell^2 + 2 \kappa C_\text{adj} \qquad \qquad
   2 \mathfrak{C} = \kappa C_\rho^2 + C_\rho
\end{equation}
Taking into account that
\begin{equation}
   \frac{i_\rho \ell^2 }{C_\rho} = \frac{N}{D} \qquad C_\text{adj} = + \ell^2 h^\vee
\end{equation}
and using \eqref{eq:firstsuper} we obtain the following constraint
equation for super quasiconformal algebras, whose grade zero
algebras are different from $\mathfrak{so}(N)$:

\begin{equation}
   \label{eq:cond1f}
   \frac{h^\vee}{i_\rho} = 5 + \frac{3D}{N(N-1)}\left( N - 8 \right) .
\end{equation}
 Now we also have
\begin{equation}
\label{eq:FF2f}
 \left[ F, \, F^\alpha\right] = \frac{\kappa}{x^3} \left( \left( \mathcal{C}_2 + \mathfrak{C} \right) \xi^\alpha +
  \xi^\alpha  \left( \mathcal{C}_2 + \mathfrak{C} \right) + \kappa \left[ \mathcal{C}_2, \left[ \xi^\alpha, \,
  \mathcal{C}_2 \right] \right] \right)
\end{equation}
Hence the constraint imposed by the commutation relation $\left[
F,\,F^\alpha \right]=0$ is
\begin{equation}
 2 \xi^\alpha \left( \mathcal{C}_2 + \mathfrak{C} \right) - C_\rho \xi^\alpha +
    \left( 2 + 4 \kappa C_\rho \right) {\left( \lambda_a \right)^\alpha}_\beta -
    \kappa C_\rho^2 \xi^\alpha - 4 \kappa {\left( \lambda_a \lambda_b \right)^\alpha}_\beta \xi^\beta
    J^b J^a =0
\end{equation}
which,  upon contraction with $\xi^\gamma \Omega_{s\gamma\alpha}$
leads to the following condition
\begin{equation}
   \label{eq:cond2f}
   \frac{h^\vee}{i_\rho} = -1 + \frac{D}{N\left(N-1\right)}\left( N + 8 \right).
\end{equation}
These two conditions \eqref{eq:cond1f} and \eqref{eq:cond2f} agree
provided
\begin{equation}
    D = \frac{3 N \left( N - 1 \right) }{ 16 - N }
\end{equation}
which is also the condition for them to agree with the equation
\begin{equation}
h^\vee = 2 i_\rho \left ( \frac{D}{N} + \frac{3D}{N(1-N)} +1 \right)
\end{equation}
which was obtained as a consistency condition for the existence of
certain class of infinite dimensional nonlinear superconformal
algebras \cite{bg97,fradline}.

Looking at the tables  of simple Lie superalgebras
\cite{bg97,fradline} consistent with our Ansatz  we find the
following simple Lie algebras $\mathfrak{g}_0$ and their irreps of
dimension $N$ that satisfy these conditions:
\begin{equation}
  \begin{array}{ccc}
     \mathfrak{g}_0 &  D & N \\
    \hline
     \mathfrak{b}_3 & 21 & 8_s \\
     \mathfrak{g}_2 & 14 &  7
  \end{array}
\end{equation}
These solutions correspond to the Lie superalgebras
$\mathfrak{f}(4)$ with even subalgebra $\mathfrak{b}_3 \oplus
\mathfrak{sl}(2,\mathbb{R})$ and $\mathfrak{g}(3)$ with even
subalgebras $\mathfrak{g}_2 \oplus \mathfrak{sl}(2,\mathbb{R}) $.

We should note that the real forms with an even subgroup of the form
$H\times SL(2,\mathbb{R})$, with $H$ simple,  admit unitary
representations only if $H$ is compact.
\section{Minimal representations of $OSp\left(N \vert 2,\mathbb{R}\right)$ \\ and
$D(2,1;\alpha)$ }
\subsection{$OSp(N\vert 2,\mathbb{R})$}
For the Lie superalgebras $\mathfrak{osp}(N\vert2,\mathbb{R})$ the
constraint equation \eqref{eq:firstsuper} does not follow from the
commutation relations and hence must not be imposed. In this case
the bilinears $ \left( \xi^\beta \xi^\alpha - \xi^\alpha \xi^\beta \right)$
generate the Lie algebra of the
even subgroup $SO(N)$ ( super analog of $Sp(2n,\mathbb{R})$. This realization corresponds to the
singleton representation of $SO(N)$ and its quadratic Casimir is
a \emph{c}-number. As a consequence of this the Jacobi identities
require either that we set $\kappa=0$ as in the case of
the minimal realization of $Sp(2n+2,\mathbb{R})$. Hence the
minimal realization reduces to a realization in terms of
bilinears of fermionic and bosonic oscillators. Thus the minimal
unitary representations of $OSp(N\vert2,\mathbb{R})$ are simply the
supersingleton representations. The singleton supermultiplets of
$OSp(N\vert2,\mathbb{R})$ were studied in \cite{GST}.

\subsection{$D(2,1;\sigma)$}
There is a one parameter family of simple Lie superalgebras of the
same dimension that has no analog in the theory of ordinary Lie
algebras. It is the family $D\left(2,1;\sigma\right)$. The real
forms of interest to us that admit unitary representations has the
even subgroup $SU(2)\times SU(2) \times SL(2,\mathbb{R})$. It has a
five grading of the form
\begin{equation}
D\left(2,1;\sigma\right) = \mathbf{1} \oplus \left(
\mathbf{2},\mathbf{2} \right) \oplus
  \left( \mathfrak{su}\left(2\right) \oplus \mathfrak{su}\left(2\right) \oplus \Delta \right)
  \oplus \left( \mathbf{2},\mathbf{2} \right) \oplus  \mathbf{1}
\end{equation}
Let $X^{\alpha, \Dot{\alpha}}$ be 4 fermionic oscillators with
canonical anti-commutation relations:
\begin{equation}
   \left\{ X^{\alpha,\Dot{\alpha}}, X^{\beta,\Dot{\beta}} \right\} = \epsilon^{\alpha\beta} \epsilon^{\Dot{\alpha}\Dot{\beta}}
\end{equation}
where $\alpha, \Dot{\alpha},..$ denote the spinor indices of the two
$SU(2)$ factors.  Let
\begin{equation}
  E = \frac{1}{2} x^2 \qquad E^{\alpha,\Dot{\alpha}} = x X^{\alpha,\Dot{\alpha}} \qquad
   \Delta = \frac{1}{2} \left(x p + p x \right)
\end{equation}
and take the generators of $\mathfrak{g}^0$ to be of the form
\begin{equation}
\begin{split}
   M_{(1)}^{\alpha,\beta} &= \frac{1}{4} \epsilon_{\Dot{\alpha}\Dot{\beta}}
    \left( X^{\alpha,\Dot{\alpha}} X^{\beta,\Dot{\beta}} + X^{\beta,\Dot{\beta}} X^{\alpha,\Dot{\alpha}}  \right) \\
   M_{(2)}^{\Dot{\alpha},\Dot{\beta}} &= \frac{1}{4} \epsilon_{\alpha\beta}
    \left( X^{\alpha,\Dot{\alpha}} X^{\beta,\Dot{\beta}} + X^{\beta,\Dot{\beta}} X^{\alpha,\Dot{\alpha}}  \right)
\end{split}
\end{equation}
They satisfy commutation relations of $\mathfrak{su}(2) \oplus
\mathfrak{su}(2)$
\begin{equation}
\begin{split}
   \left[ M^{\alpha,\beta}_{(1)}, M^{\lambda,\mu}_{(1)} \right] &=
       \epsilon^{\lambda\beta} M^{\alpha,\mu}_{(1)} + \epsilon^{\mu\alpha} M^{\beta,\lambda}_{(1)} \\
   \left[ M^{\Dot{\alpha},\Dot{\beta}}_{(2)}, M^{\Dot{\lambda},\Dot{\mu}}_{(2)} \right] &=
       \epsilon^{\Dot{\lambda}\Dot{\beta}} M^{\Dot{\alpha},\Dot{\mu}}_{(2)} +
       \epsilon^{\Dot{\mu}\Dot{\alpha}} M^{\Dot{\beta},\Dot{\lambda}}_{(2)} \\
   \left[ M^{\alpha,\beta}_{(1)}, M^{\Dot{\lambda},\Dot{\mu}}_{(2)} \right] &= 0
\end{split}
\end{equation}
Their quadratic Casimirs are
\begin{equation}
   \mathcal{I}_4 = \epsilon_{\alpha\beta}\epsilon_{\lambda\mu} M_{(1)}^{\alpha\lambda} M_{(1)}^{\beta\mu}
  \qquad
   \mathcal{J}_4 = \epsilon_{\Dot{\alpha}\Dot{\beta}}\epsilon_{\Dot{\lambda}\Dot{\mu}}
         M_{(2)}^{\Dot{\alpha}\Dot{\lambda}} M_{(2)}^{\Dot{\beta}\Dot{\mu}}
\end{equation}
Their sum is a c-number $\mathcal{I}_4 + \mathcal{J}_4 =
-\frac{3}{2}$. We use just one to construct generator of
$\mathfrak{g}^{+2}$
\begin{equation}
   F = \frac{1}{2} p^2 + \frac{\sigma}{x^2} \left( \mathcal{I}_4 + \frac{3}{4} + \frac{9}{8} \sigma \right)
\end{equation}
and
\begin{equation}
   F^{\alpha\Dot{\alpha}} = -i \left[ E^{\alpha\Dot{\alpha}}, F \right]
\end{equation}
Then
\begin{equation}
   \left[ F^{\alpha\Dot{\alpha}}, F^{\beta\Dot{\beta}} \right] = 2 \epsilon^{\alpha\beta} \epsilon^{\Dot{\alpha}\Dot{\beta}} F
  \qquad \left[ F^{\alpha\Dot{\alpha}}, F \right ] = 0
\end{equation}
Also
\begin{equation}
  \left[ F^{\alpha\Dot{\alpha}}, E^{\beta\Dot{\beta}} \right] =
       \epsilon^{\alpha\beta} \epsilon^{\Dot{\alpha}\Dot{\beta}} \Delta -
       \left( 1- 3 \sigma \right) i \epsilon^{\alpha\beta} M_{(2)}^{\Dot{\alpha}\Dot{\beta}} -
       \left( 1+ 3 \sigma \right) i \epsilon^{\Dot{\alpha}\Dot{\beta}} M_{(1)}^{\alpha\beta}
\end{equation}
The parameter $\sigma$ is left undetermined by the Jacobi
identities. For $\sigma=0$ the superalgebra
$D\left(2,1,\sigma\right)$ is isomorphic to $OSp(4 \vert
2,\mathbb{R})$ and for the values $\sigma= \pm \frac{1}{3}$ it
reduces to
\begin{equation*}
SU\left(2 \vert 1,1\right)\times SU\left(2\right)
\end{equation*}

 {\bf Acknowledgement:} One of us (M.G.) would like to thank the organizers of the
 workshop on ``Mathematical Structures in String Theory'' at KITP where part of this work was done.
 We would  like to thank Andy Neitzke, Boris Pioline, David Vogan and Andrew Waldron for stimulating
  discussions and correspondence.
 This work was
supported in part by the National Science Foundation under grant
number PHY-0245337. Any opinions, findings and conclusions or
recommendations expressed in this material are those of the authors
and do not necessarily reflect the views of the National Science
Foundation. We would also like to thank the referee for several
suggestions for improving the presentation of the results.

\end{document}